\documentclass[fleqn,usenatbib]{mnras}

\usepackage{newtxtext,newtxmath}

\usepackage[T1]{fontenc}

\DeclareRobustCommand{\VAN}[3]{#2}
\let\VANthebibliography\thebibliography
\def\thebibliography{\DeclareRobustCommand{\VAN}[3]{##3}\VANthebibliography}

\usepackage{graphicx}	
\usepackage{amsmath}	
\newsavebox{\measurebox}
\usepackage{subcaption}
\usepackage{color}
\usepackage{ulem}
\usepackage{adjustbox}
\usepackage{tablefootnote}
\usepackage{threeparttable}
\usepackage[justification=centering]{caption}
\usepackage{booktabs}
\usepackage{tabularx}
\usepackage{soul}
\usepackage{gensymb}
\usepackage[dvipsnames]{xcolor}

\title[Extended radio emission in HCG~97b]{
Ram-pressure stripped radio tail and two ULXs in the spiral galaxy HCG~97b}

\author[Dan Hu et al.]
{Dan Hu,$^{1}$\thanks{E-mail: hudan.bazhaoyu@mail.muni.cz}
Michal Zajaček,$^{1}$
Norbert Werner,$^{1}$
Romana Grossová,$^{2,1}$
Pavel Jáchym,$^{2}$
\newauthor
Ian D. Roberts,$^{3}$
Alessandro Ignesti,$^{4}$
Jeffrey D.P. Kenney,$^{5}$
Tomáš Plšek,$^{1}$
Jean-Paul Breuer,$^{1}$
\newauthor
Timothy Shimwell,$^{6,3}$
Cyril Tasse,$^{7,8}$
Zhenghao Zhu,$^{9}$
and 
Linhui Wu$^{9}$
\\
$^{1}$Department of Theoretical Physics and Astrophysics, Faculty of Science, Masaryk University, Kotl\'{a}\v{r}sk\'{a} 2, Brno, 611 37, Czech Republic\\
$^{2}$Astronomical Institute of the Czech Academy of Sciences, Bo\v{c}n\'{i} II 1401, Prague, 141 00, Czech Republic\\
$^{3}$Leiden Observatory, Leiden University, PO Box 9513, 2300 RA Leiden, The Netherlands\\ 
$^{4}$INAF—Astronomical Observatory of Padova, vicolo dell’Osservatorio 5, I-35122 Padova, Italy\\
$^{5}$Department of Astronomy, Yale University, New Haven, CT 06511, USA\\
$^{6}$ASTRON, The Netherlands Institute for Radio Astronomy, Postbus 2, 7990 AA Dwingeloo, The Netherlands\\
$^{7}$GEPI \& ORN, Observatoire de Paris, Université PSL, CNRS, 5 Place Jules Janssen, 92190 Meudon, France\\
$^{8}$Department of Physics \& Electronics, Rhodes University, PO Box 94, Grahamstown, 6140, South Africa\\
$^{9}$Shanghai Astronomical Observatory, Chinese Academy of Sciences, Nandan Road 80, Shanghai, China
}

\date{Accepted 2003 October 17. Received 2023 October 16; in original form 2023 April 24}

\pubyear{2023}

\begin{document}
\label{firstpage}
\pagerange{\pageref{firstpage}--\pageref{lastpage}}
\maketitle

\begin{abstract}
We report LOFAR and VLA detections of extended radio emission in the spiral galaxy HCG~97b, hosted by an X-ray bright galaxy group. The extended radio emission detected at 144~MHz, 1.4~GHz and 4.86~GHz is elongated along the optical disk and has a tail that extends 27~kpc in projection towards the centre of the group at GHz frequencies or 60~kpc at 144~MHz.
Chandra X-ray data show two off-nuclear ultra-luminous X-ray sources (ULXs), with the farther one being a plausible candidate for an accreting intermediate-mass black hole (IMBH).
The asymmetry observed in both CO emission morphology and kinematics indicates that HCG~97b is undergoing ram-pressure stripping, with the leading side at the southeastern edge of the disk. 
Moreover, the VLA 4.86 GHz image reveals two bright radio blobs near one ULX, aligning with the disk and tail, respectively. The spectral indices in the disk and tail are comparable and flat ($\alpha > -1$), suggesting the presence of recent outflows potentially linked to ULX feedback. This hypothesis gains support from estimates showing that the bulk velocity of the relativistic electrons needed for transport from the disk to the tail is approximately $\sim 1300$~$\rm km~s^{-1}$. This velocity is much higher than those observed in ram-pressure stripped galaxies ($100-600$~$\rm km~s^{-1}$), implying an alternative mechanism aiding the stripping process.
Therefore, we conclude that HCG~97b is subject to ram pressure, with the formation of its stripped radio tail likely influenced by the putative IMBH activities. 
\end{abstract}



\section{Introduction}

The properties of galaxies within a dense environment are influenced via various mechanisms, such as mergers \citep[e.g.,][]{SB51,MH94a,MH94b,springel00,misquitta2023}, tidal interactions \citep[e.g.,][]{BV90,valluri93,mayer06,chung07,lokas20},  ram-pressure stripping \citep[e.g.,][]{GG72,QMB00,jachym14,roberts21a}, and viscous stripping \citep[e.g.,][]{nulsen82,QMB00,roediger15}. Mergers and tidal interactions are primarily gravitational effects that affect both stellar and gas components of a galaxy, while the other two are predominantly hydrodynamical effects that only influence the gas content of the galaxy. 
Ram-pressure stripping can effectively remove the interstellar medium (ISM) from galaxies that are falling into cluster environments, leading to the subsequent formation of gas tails. The most extreme examples of galaxies undergoing intense ram-pressure stripping are so-called jellyfish galaxies (e.g., \citealt{smith10,ebeling14,fumagalli14}; and see \citealt{BFS22}, for a review).

Extended, one-sided gas tails from ram-pressure stripped galaxies were observed in some galaxy clusters via $\rm H\alpha$ \citep[e.g.,][]{gavazzi01,sun07,gavazzi17}, H\textsc{i} \citep[e.g.,][]{kenney14,deb20}, CO \citep[e.g.,][]{jachym14,jachym19}, X-ray emission \citep[e.g.,][]{sun06,sun10,poggianti19}, and radio continuum emission \citep{GJ87,murphy09,vollmer09,vollmer13,roberts22,muller21,ignesti22}. 
In general, radio continuum emission in ram-pressure stripped galaxies is mainly caused by non-thermal synchrotron emission of relativistic cosmic ray electrons (CRe), which are accelerated by supernovae (SNe) shocks (e.g., \citealt{condon92}, for a review). These relativistic CRe are then stripped from the galaxy by ram pressure due to the relative motion between the galaxy and the ambient environment and possibly further re-accelerated by turbulence, intracluster medium (ICM) shocks, or new SNe. As such, radio continuum emission presents a valuable complementary tool for studying the phenomenon of ram-pressure stripped tails.

The process of ram-pressure stripping is generally observed in galaxy clusters due to their high ICM densities and large infall velocities. 
However, the ram-pressure exerted by the intragroup medium (IGrM) is commonly considered to be inefficient, resulting in a stripping process extended over a longer timescale ($\sim 3$~Gyr), as supported by cosmological hydrodynamical simulation performed by \citet{oman21}.
Despite this, evidence of ram-pressure stripping in galaxy groups is still observed \citep[e.g.,][]{davis97,SSC04,machacek05,kantharia05,rasmussen06}.
Recently, the number of known ram-pressure stripped galaxies in galaxy groups was increased as \citet{roberts21b} identified 60 jellyfish galaxies with extended, asymmetric radio continuum tails in low-mass systems with the LOFAR Two-metre Sky Survey (LoTSS; \citealt{shimwell17,shimwell22}), that has a high sensitivity ($\sim 0.1$~$\rm mJy~beam^{-1}$) at low frequency (144~MHz). 
However, it should be noted that ram-pressure stripping might not be the only mechanism explaining the gas morphology seen in these low-mass systems, which might also be affected by tidal interaction as suggested by the aforementioned authors or by AGN activity \citep[e.g.,][]{poggianti17}.

To further investigate the gas removal from galaxies in the group environment, we conducted a multi-wavelength study of the spiral galaxy HCG~97b (alternative name IC5359), which resides at $\sim 100$~kpc southeast of the central galaxy HCG~97a (IC5357) of the galaxy group HCG~97. This galaxy is moderately inclined, appearing close to edge-on but not completely perpendicular to our line of sight. The bulge-to-total luminosity ratio is $\sim 0.08$ using the SDSS g-band data \citep{bizyaev14}. 
The mean Heliocentric radial velocity of HCG~97b is 6940$\pm 74 \rm~km s^{-1}$ \citep{hickson92}, and the stellar mass of HCG~97b is $\sim 1.7 \times 10^{10}$~$\rm M_{\sun}$ estimated by using $K$s-band luminosity \citep{bitsakis11}. The apparent diameter $\rm D_{25}$ (defined by the isophote at the brightness of 25 $\rm mag~arcsec^{-2}$ in the $B$ band) of the spiral galaxy HCG~97b is $1.23\pm 0.11$~arcmin (from HyperLeda database\footnote{\url{http://leda.univ-lyon1.fr}}; \citealt{makarov14}), corresponding to $35 \pm 3.1$~kpc ($1\arcmin \approx 27$~kpc), implying it is a large spiral galaxy comparable to our Milky Way ($\rm D_{25} = 26.8 \pm 1.1$~kpc; \citealt{goodwin98}). 
The results of this work suggest the presence of two off-nuclear X-ray sources that are likely ULXs, as well as an extended, one-sided radio tail that is being stripped by ram pressure resulting from the galaxy-IGrM interaction. One of the ULXs may be a candidate for an activated IMBH and has a potential link with the nearby radio tail. Therefore, HCG~97b presents an ideal case for studying ram-pressure stripped galaxies in the group environment and investigating the potential contribution of ULXs/IMBH to the enhancement of ram-pressure stripping.

Throughout this paper, we assume the standard flat $\Lambda$CDM cosmology with parameters $H_0=70$~$\rm km~s^{-1}~Mpc^{-1}$ and $\Omega_{m} = 1 - \Omega_{\Lambda}$=0.27. At the redshift of HCG~97b ($z=0.022$), these values result in a scale of $\sim 27$~$\rm kpc~arcmin^{-1}$ and a luminosity distance of 97~Mpc. For the radio spectral index $\alpha$, we use the convention $S_{\nu} \propto \nu^{\alpha}$, where $S_{\nu}$ is the flux density at the frequency of $\nu$.

\section{Observations and data reduction}
\label{sec:sect2}
All data were retrieved from public data archives to study the X-ray emission, radio continuum emission, and molecular gas of the spiral galaxy HCG~97b.

 \subsection{Chandra observations}
\label{sec:chandra} 
Chandra data analyzed in this work were obtained using the S3 chip of the Advanced CCD Imaging Spectrometer (ACIS) in the VFAINT mode on January 14, 2005 (ObsID 4988; 57.4~ks). We followed the standard procedure suggested by the Chandra X-ray Center by using the Chandra Interactive Analysis of Observations (\textsc{ciao}) v4.15.1 and Chandra Calibration Database (\textsc{caldb}) v4.10.2 to carry out data reduction. The detailed data reduction and analysis descriptions are provided in \citet{hu21}.
In this work, the spectra were extracted using the \textsc{ciao} tool \texttt{specextract} and fitted using \textsc{xspec} v12.13.0 and \textsc{atomdb} v3.0.9.

\subsection{LOFAR observations}
\label{sec:lofar} 
HCG~97 is within the field-of-view of three pointings (with identifiers of P354+01, P357+01, and ZwCL2341.1+0000) from the LOFAR Two-metre Sky Survey (LoTSS; \citealt{shimwell17}), a deep survey with a typical sensitivity of $0.1~\rm mJy~beam^{-1}$ at good elevations and a resolution of $6\arcsec$ at the central frequency of 144~MHz. In all three 8~hrs duration pointings, the target HCG~97 is significantly offset from the pointing centre ($> 1\deg$). To combine the data and improve upon the standard pipeline-processed image quality, we extracted and re-calibrated the LOFAR data following the procedure outlined in \citet{vanweeren21}. To do this we first defined a target region with a radius of $\sim 0.6\deg$ to include both HCG~97 and enough bright sources for successful re-calibration.
Then, sources outside of this region were removed using the full direction-dependent calibration solutions (using kMs and DDFacet; \citealt{tasse14,tasse18}) before the phase centres of visibilities were shifted to the centre of the target region. Before combining the three sets of visibility data, we carried out a beam response correction and updated the visibility weights accordingly. Finally, the re-calibrated step was performed with three rounds of "tecandphase" calibration and gain calibration using the Default Pre-Processing Pipeline (\texttt{DPPP}; \citealt{vandiepen18}). 
The final image was produced with \texttt{WSClean} \citep{offringa14} with a Briggs robust \citep{briggs95} parameter of $0.0$. The rms of the final image is $0.22~\rm mJy~beam^{-1}$ with a beam size of $16.90\arcsec \times 10.60\arcsec$ (see Table~\ref{tab:radio-data-prop}). Due to projection effects and the low declination of our target, the sensitivity is lower than is achieved for higher elevation LoTSS fields but is consistent with other low declination LOFAR studies \citep[e.g.,][]{hale19}.

\subsection{VLA observations}
\label{sec:vla} 
Radio continuum data of the galaxy group HCG~97 at GHz frequencies were retrieved from the Very Large Array (VLA) archives. Two VLA observations were taken on November 13, 2006 and April 7, 2007 (Project ID: AY171) in the C- and D-configuration with central frequencies of 1.4 GHz and 4.86 GHz, respectively. The total observation time for each configuration is around 2.35 hours. 
The pre-upgrade VLA data were analyzed and calibrated by using the Common Astronomy Software Applications (\textsc{casa}) package \citep{mcmullin07} basically following the official "Jupiter continuum calibration tutorial\footnote{\url{https://casaguides.nrao.edu/index.php/Jupiter:\_continuum\_polarization\_calibration}}" provided by NRAO. In brief, we applied the automatic flagging algorithm \texttt{tfcrop} to identify and remove the radio frequency interference (RFI), and used 0137+331 (3C48) as the flux calibrator and 0059+001 as the phase calibrator to determine the gain solutions for the target (see \citealt{grossova22}, for more detail). Subsequently, further flagging and calibration steps were applied to enhance the quality of the images. After an initial imaging, at least one round of self-calibration was executed to refine the calibration. We used the \textsc{casa} task \texttt{tclean} with the gridding algorithm \texttt{wproject} and \texttt{multiscale} clean techniques to address widefield non-coplanar baseline effect and ensure proper imaging of the extended radio emission. The target was finally imaged with a Briggs robust 0.0 weighting after the primary-beam correction.
The resulting images reach an rms noise of 0.045~$\rm mJy~beam^{-1}$ and 0.028~$\rm \mu Jy~beam^{-1}$ and the synthesized beam sizes of $16.75\arcsec \times 14.46\arcsec$ and $15.63\arcsec \times 13.50\arcsec$ at 1.4 GHz and 4.86 GHz (see Table~\ref{tab:radio-data-prop}), respectively.

\begin{table}
 \caption{Properties of LOFAR and VLA data.}
 \label{tab:radio-data-prop}
 \centering
 \renewcommand{\arraystretch}{1.3}
 \begin{threeparttable}
  \begin{tabularx}{\columnwidth}{lcccc}
  \hline
  Telescope & $\nu_{\rm c}$ & $\sigma$ & PSF & $\sigma'\tnote{a} $  \\
    & (MHz) &  ($\rm mJy~beam^{-1}$) & ($\arcsec \times \arcsec$) &  ($\rm mJy~beam^{-1}$) \\
  \hline
  LOFAR &  144  & 0.22  &  $16.90 \times 10.60$  & 0.47 \\
  \hline
  VLA   &  1400  &  0.045 &  $16.75 \times 14.46$  & 0.050 \\
        &  4860  &  0.028 &  $15.63 \times 13.50$  & 0.038 \\
  \hline
  \end{tabularx}
  \begin{tablenotes}
    \item [a] The measurements of rms from new images after imaging with new parameters (see details in Subsection~\ref{sect:radio_spc}) and smoothing with a $20\arcsec \times 20\arcsec$ Gaussian beam.
  \end{tablenotes}
 \end{threeparttable}
\end{table}

\subsection{ALMA observations}
\label{sec:alma} 
The molecular disk traced by the CO (2-1) line transition of the spiral galaxy HCG~97b was observed by the 12m Atacama Large Millimeter/submillimeter Array (ALMA) and 7m Atacama Compact Array (ACA) (Project ID: 2018.1.00657.S). Using the Band 6 receiver, a spectral window centred around the CO (2-1) line at 225.262~GHz. 
The data were calibrated using the ALMA calibration pipeline (version 42030M) in \textsc{casa} (version 5.4.0-68) and then imaged with the Briggs weighting (robust=0.5) to balance the sensitivity and resolution. The spectral resolution is 8~MHz, corresponding to 10~$\rm km~s^{-1}$.
The resulting synthesized beams in 12m ALMA and 7m ACA image data cubes are $0.53\arcsec \times 0.42\arcsec$ and $7.73\arcsec \times 4.36\arcsec$, respectively. Correspondingly, the achieved rms noises for these data cubes are 3~$\rm mJy~beam^{-1}$ and 9~$\rm mJy~beam^{-1}$, respectively.

\section{Results}
\label{sec:sect3}

\subsection{X-ray properties of two X-ray sources in HCG~97b}
\label{sect:xray} 

\begin{figure*}
    \centering
    \includegraphics[scale=.34]{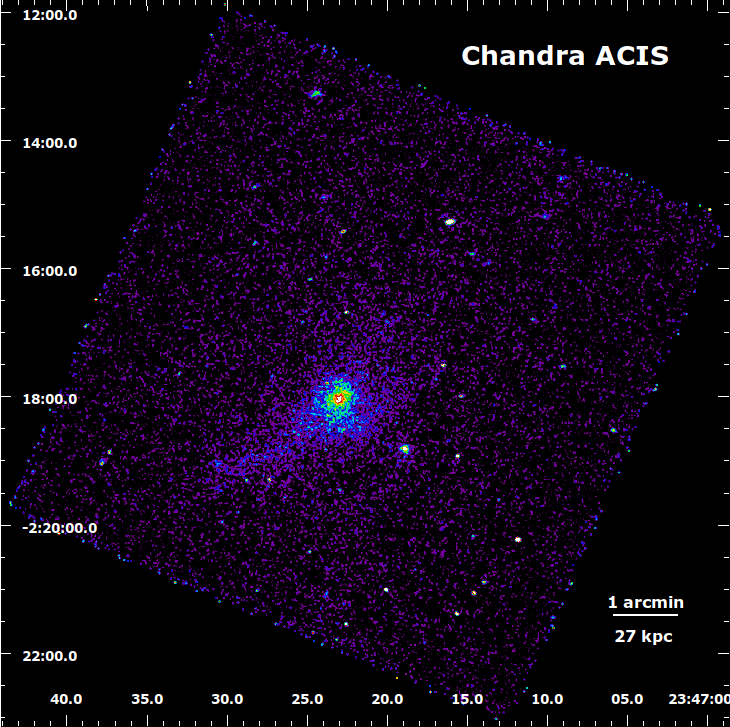}
    \includegraphics[scale=.34]{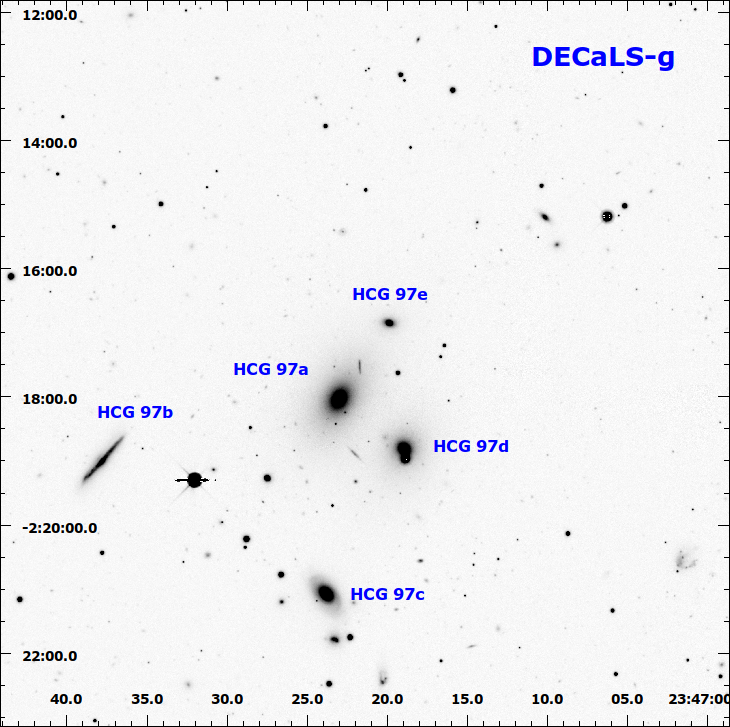}
    \includegraphics[scale=.35]{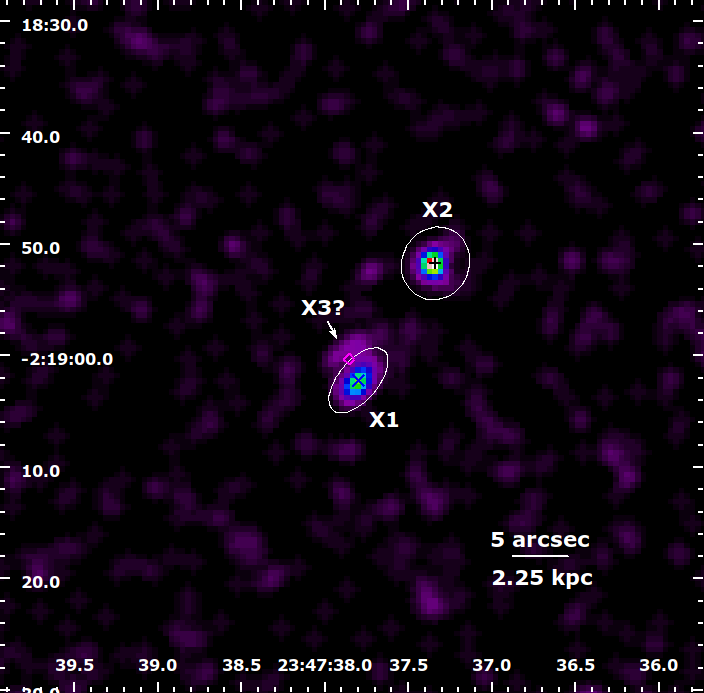}
    \includegraphics[scale=.35]{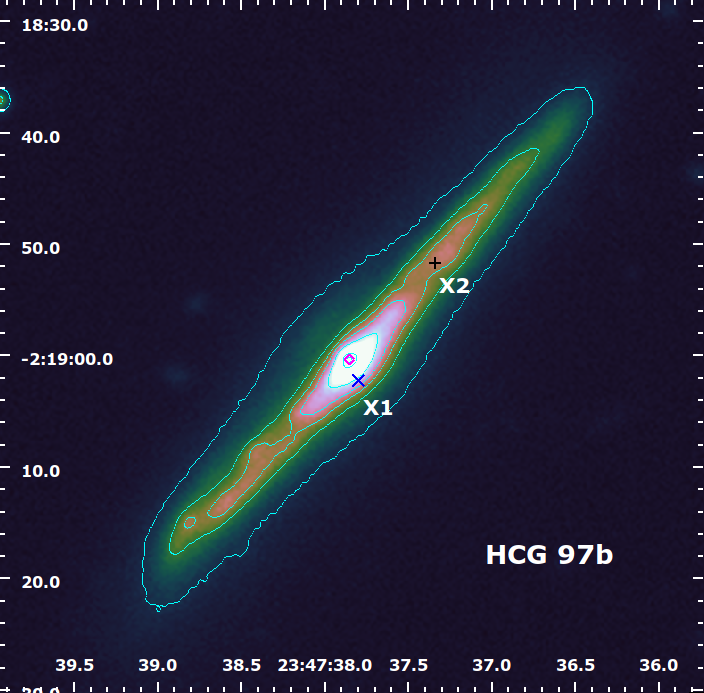}
    \caption{Top: Exposure-corrected $0.5-7$~keV Chandra ACIS image of HCG~97 (left) and the DECalS g-band optical image (right). Five member galaxies of HCG~97 are marked. Bottom: Zoom-in view of HCG~97b on the Chandra image (left) and DECalS g-band image with optical contours (right). \texttt{wavdetect} regions are presented with two ellipses. For source X1, the major- and minor-axis diameters are $\sim 7.0\arcsec$ and $3.6\arcsec$, respectively; for X2, the major- and minor-axis diameters are $\sim 6.6\arcsec$ and $6.0\arcsec$, respectively. X-ray sources X1 and X2 are marked with a cross and plus sign, respectively. The optical centre is marked with a diamond. The label X3 indicates an extended X-ray source candidate potentially related to the source X1.}
    \label{fig:overall}
\end{figure*}

The exposure-corrected $0.5-7$~keV Chandra ACIS image and DECaLS \citep{dey19} g-band optical image of HCG~97 are presented in Figure~\ref{fig:overall}, as well as the zoom-in images of the spiral galaxy HCG~97b. X-ray emission of the group shows a clear X-ray plume extending southeast from the group centre, indicating a disturbed galaxy group. At the location of the spiral galaxy HCG~97b, there are two X-ray sources (X1 and X2). These two X-ray sources were also identified by \texttt{wavdetect} with the \texttt{scale} parameters of 1.0, 2.0, 4.0, 8.0, 16.0 and the significance threshold (\texttt{sigthresh}) of $10^{6}$. The X2 is point-like whilst X1 has an elongated shape with major- and minor-axis diameters of $\sim 7.0 \arcsec$ and $3.6\arcsec$, respectively. Neither X1 nor X2 is the central AGN as they are off-nuclear and have projected separations of $\sim 0.86$~kpc ($\sim 1.9\arcsec$) and $\sim 5.2$~kpc ($\sim 11.6\arcsec$), respectively, from the optical centre of the galaxy (marked as a red diamond in Figure~\ref{fig:overall}). 

\begin{figure}
    \centering
    \includegraphics[scale=.3]{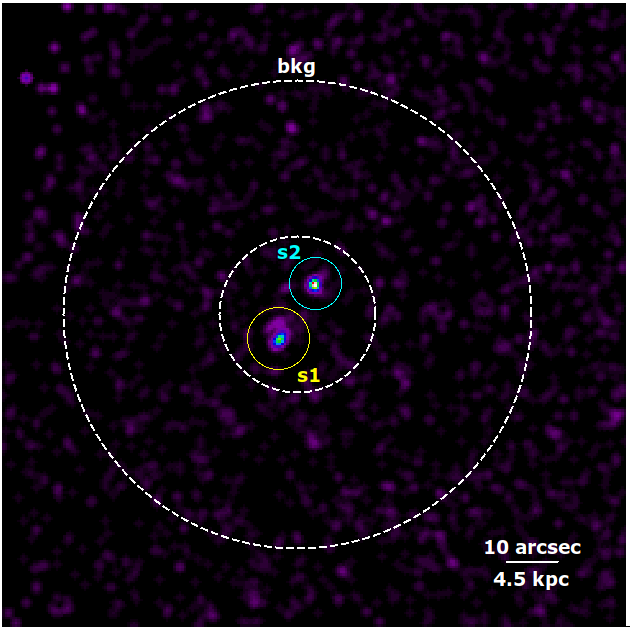}
    \caption{Regions used to extract spectra are presented in the zoom-in Chandra image.}
    \label{fig:xray-spc}
\end{figure}

In order to investigate the nature of these two X-ray sources, we extracted the spectra from circular regions (s1 and s2; see Figure~\ref{fig:xray-spc}) with radii of 6$\arcsec$ and 5$\arcsec$ for sources X1 and X2, respectively. Concerning the possible contamination from group emission, the background spectrum was extracted from an annular region spanning from $15\arcsec$ to $45\arcsec$ (bkg; see Figure~\ref{fig:xray-spc}) and encircling two point sources. To model the spectra of two X-ray sources, we used an absorbed power-law model (\texttt{phabs*phabs*powerlaw}): the first absorption component was set equal to the line-of-sight Galactic absorption $N_{\rm H}=3.4 \times 10^{20}$~$\rm cm^{-2}$ \citep{hi4pi16}, while the other one was set to be free to account for the possible intrinsic absorption of the source. The spectra were fitted and analyzed using the Bayesian X-ray Analysis (BXA; \citealt{buchner14}) software and C-statistic to acquire well-defined parameter constraints for spectra with a low number of counts. 

For the source X1, the best-fit photon index is $\Gamma=2.65\pm 0.87$ and the absorption column density is $N_{\rm H}=0.42 \pm 0.26 \times 10^{22}$~$\rm cm^{-2}$. While for source X2, because of the low number of counts in the region s2, the parameter of the photon index cannot be well constrained. Therefore, we fixed the photon index at 1.6 \footnote{We have tested different indices, i.e., from 1.5 to 2.0, the variation in the $N_{\rm H}$ was within 20\%, which still suggests source X2 is an obscured source.} and obtained $N_{\rm H}= 2.28 \pm 0.57 \times 10^{22}$~$\rm cm^{-2}$, indicating a mildly obscured source. This is consistent with source X2 being surrounded by molecular gas that can be seen in the CO distributions (see Subsection~\ref{sect:CO}).
The absorbed $2-10$~keV X-ray luminosities of X1 and X2 are $L_{\rm X1,~2-10~keV}= 3.78 \times 10^{39}$~$\rm erg~s^{-1}$ and $L_{\rm X2,~ 2-10~keV}= 1.80\times 10^{40}$~$\rm erg~s^{-1}$, respectively. The derived X-ray luminosities indicate that the off-centre X-ray sources X1 and X2 could be ULXs ($L_{\rm X}> 10^{39}$~$\rm erg~s^{-1}$; \citealt{kaaret17}). \citet{wang16} supports this classification as these two sources (names IC5359-X1 and IC5359-X2) were listed as ULXs in their X-ray point source catalogue. 

\begin{figure}
    \centering
    \includegraphics[scale=.55]{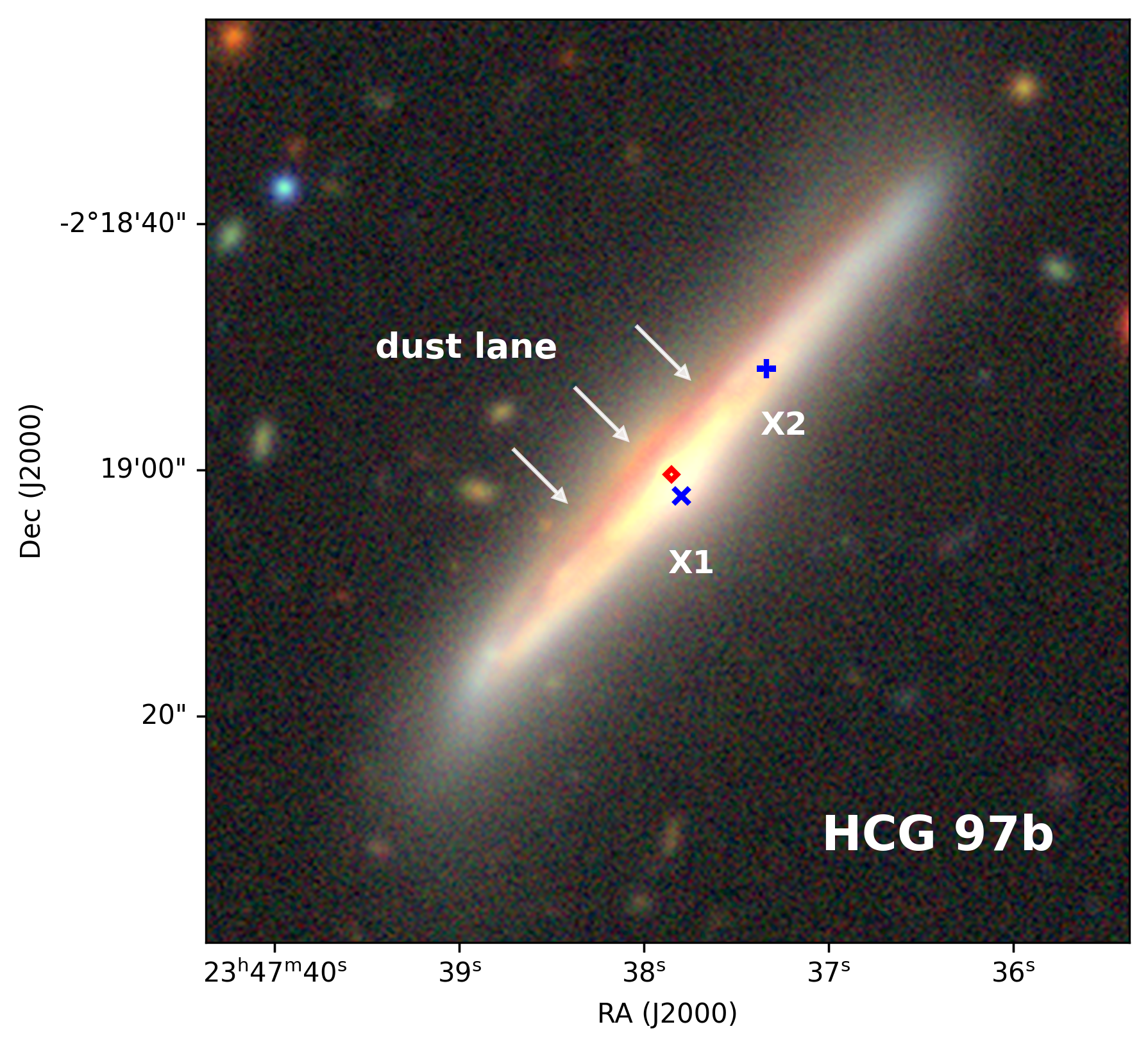}
    \caption{Composite DECalS optical image using g-, r- and z-band data. X-ray sources X1 and X2 are marked with a cross and plus sign, respectively. The optical centre is marked with a red diamond. The dust lane is also labelled.}
    \label{fig:rgb-decals}
\end{figure}

Near X1, we notice a faint, small-scale X-ray excess, which could be a potential X-ray source (X3; see Figure~\ref{fig:overall}), although the \texttt{wavdetect} failed to identify it. Furthermore, the spectrum of this faint X-ray substructure could not be constrained due to the low-counts limitation. Since the X-ray sources X1 and putative-X3 reside in projection on either side of the optical galaxy centre (see the red diamond in Figure~\ref{fig:overall}), it is possible that these are fingerprints of two star-forming regions at the two sides of the galactic bar or along it. X3 is fainter than X1, which might be because of an apparent dust lane covering X3, while X1 is not obscured by the dust (see composite DECaLS image in Figure~\ref{fig:rgb-decals}). It is also possible that sources X1 and X3 correspond to jets, but we have not yet been able to assess this with radio data at sufficient angular resolution. Future deeper X-ray data and high-resolution radio studies are needed to assess these scenarios.

\subsection{Radio properties of the spiral galaxy HCG~97b} 
\label{sect:radio}

\subsubsection{Radio image}
\label{sect:radio_image}

\begin{figure*}
    \centering
    \includegraphics[width=0.95\textwidth]{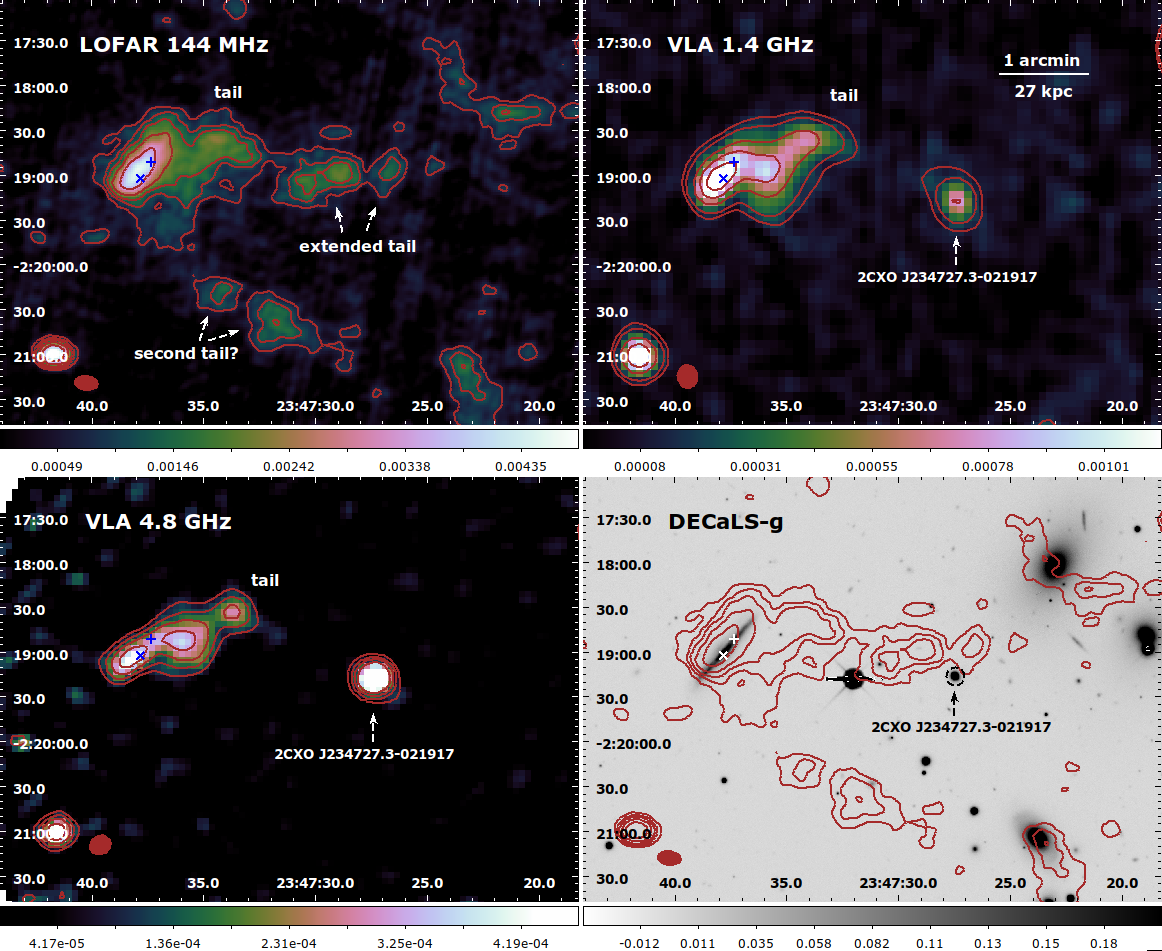}
    \caption{Top-left: LOFAR image of HCG~97b at 144~MHZ with a beam size of $16.9\arcsec \times 10.60\arcsec$. Contours show 3$\sigma$, 5$\sigma$, 7$\sigma$, 10$\sigma$ and 15$\sigma$ flux density levels of radio continuum emission (unit: $\rm Jy~beam^{-1}$). The radio tail, extended tail, and possible southern tail with low significance are all labelled. Top-right: VLA image of HCG~97b at 1.4~GHz with a beam size of $16.75\arcsec \times 14.46\arcsec$. Contours show 3$\sigma$, 5$\sigma$, 10$\sigma$, 15$\sigma$, and 20$\sigma$ flux density levels of radio continuum emission. Bottom-left: VLA image of HCG~97b at 4.86~GHz with a beam size of $15.63\arcsec \times 13.50\arcsec$. Contours show 3$\sigma$, 5$\sigma$, 7$\sigma$ and 10$\sigma$ flux density levels of radio continuum emission. Values in three radio images are in a unit of $\rm Jy~beam^{-1}$. Bottom-right: DECaLS g-band image with 144~MHz contours. Two X-ray sources, X1 and X2, are marked with a cross and plus sign, respectively. A bright point source (2CXO J234727.3-021917) was also labelled.}
    \label{fig:radio}
\end{figure*}

The radio images of HCG~97b derived from LOFAR 144~MHz, VLA 1.4~GHz and 4.86~GHz are presented in Figure~\ref{fig:radio} along with the optical image. 
All radio images reveal an asymmetric morphology, characterized by the radio luminous region located at the disk, along with distinct extraplanar radio emission. In the VLA 1.4~GHz and 4.86~GHz images, the radio emission is curved in the northwest region of the disk, roughly corresponding to the position of X2, and extends $\sim 27$~kpc ($\sim 1\arcmin$) in projection outside the disk toward the northwest. We refer to this emission as a radio tail hereafter. This radio tail was visible in \citet{bharadwaj14}, who used 1.4~GHz NRAO VLA Sky Survey (NVSS) radio contours, but this was not the focus of their study. Interestingly, we find two bright radio blobs, one above the disk, and another one outside the disk but near the position of X2 in the VLA 4.86~GHz image, potentially indicating a pair of lobes related to the ULX X2. The morphology of the radio tail is not straight and shows evidence of distortion. This might indicate a possible interaction between the radio tail and the IGrM. 

\begin{figure}
    \centering
    \includegraphics[scale=.35]{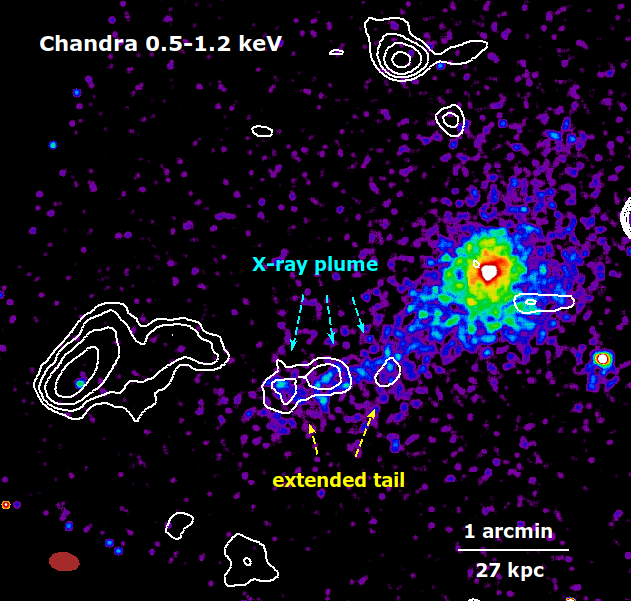}
    \caption{Exposure-corrected $0.5-1.2$~keV Chandra ACIS image of HCG~97 with LOFAR 144~MHz contours starting from $5\sigma$. }
    \label{fig:softxray-radiospc}
\end{figure}

The radio tail is also obvious at 144~MHz and is more extended than at GHz frequencies. 
At the western end of the radio tail, there is a strip of radio emission detected above $3\sigma$ in the LOFAR 144~MHz image that is possibly an outer part of the radio tail of HCG~97b (we refer to it as an extended tail hereafter), resulting in a total length of $\sim 60$~kpc at low frequency. Interestingly, as presented in Figure~\ref{fig:softxray-radiospc}, $> 5 \sigma$ contours on the LOFAR 144~MHz image show that there is a decline in brightness in the region between the radio tail and the extended tail. Furthermore, the location of the extended tail coincides with the end part of the X-ray plume from the group centre (or X-ray halo of HCG~97a; also see Figure~\ref{fig:softxray-radiospc}). 
This suggests the possibility of an interaction between the extended tail and the IGrM within the X-ray plume, which may lead to the re-ignition of aged radio plasma in the region of the extended tail, possibly due to turbulence or compression.
Additionally, we have identified two regions of diffuse radio emission with relatively low significance, situated to the southwest of HCG~97b (see Figure~\ref{fig:radio}). These regions could potentially represent a second radio tail linked to HCG~97b since there is a lack of clear optical counterparts corresponding to this radio emission, particularly for those regions where the emission exceeds an rms level of $> 5 \sigma$. To substantiate this observation, it is essential to acquire more extensive and comprehensive radio data, as well as other wavelength data (e.g., H$\alpha$ data).

\subsubsection{Radio spectrum}
\label{sect:radio_spc}

\begin{figure*}
    \centering
    \includegraphics[width=0.95\textwidth]{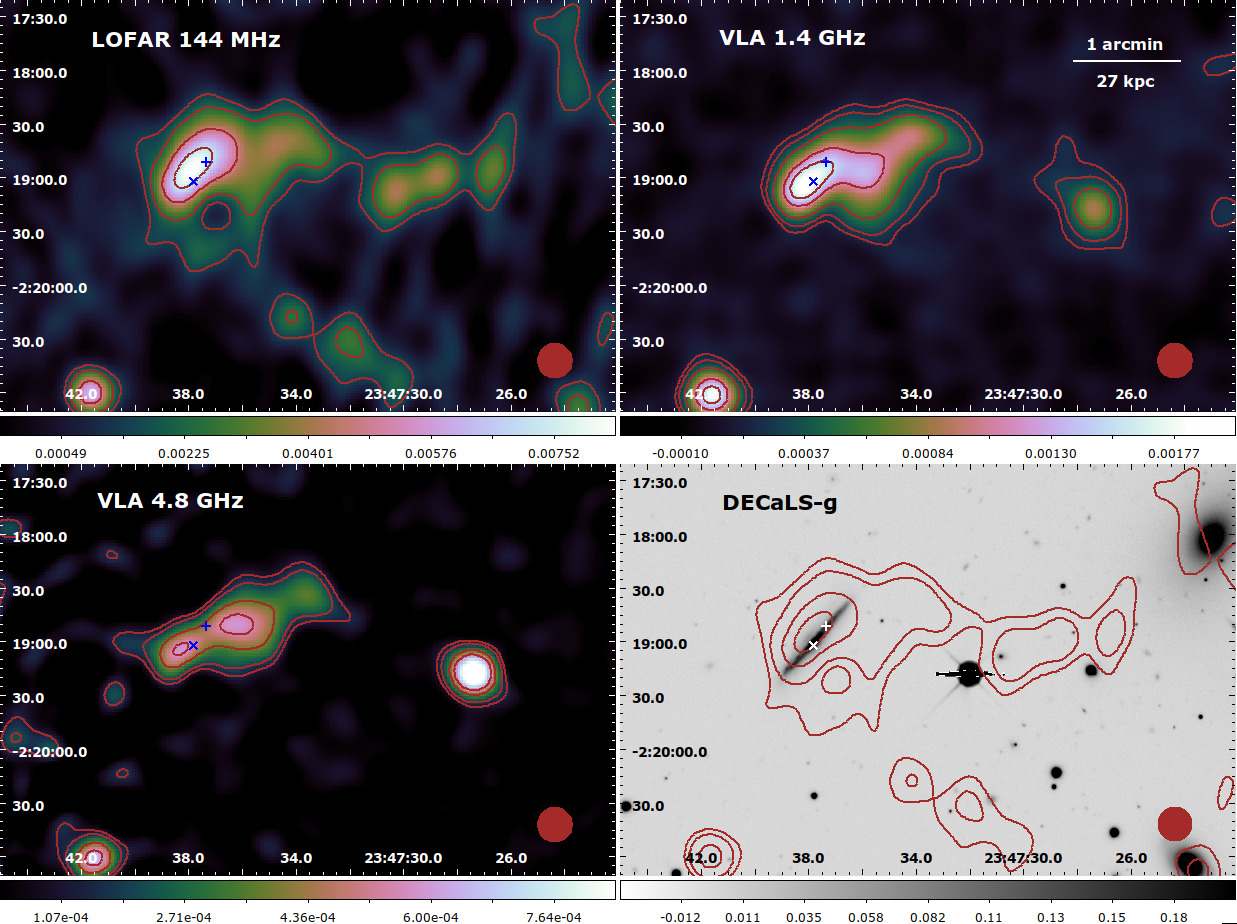}
    \caption{Same as Figure~\ref{fig:radio}, but all radio images were imaged with new parameters (see details in Subsection~\ref{sect:radio_spc}) and smoothed with a $20\arcsec \times 20\arcsec$ Gaussian beam. Contours show 3$\sigma'$, 5$\sigma'$, 10$\sigma'$, 15$\sigma'$ and 20$\sigma'$ flux density levels of radio continuum emission.}
    \label{fig:radio_smo20}
\end{figure*}

\begin{figure}
    \centering
    \includegraphics[scale=.45]{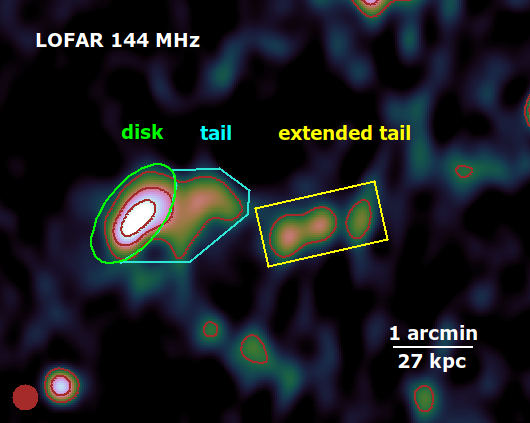}
    \caption{LOFAR 144~MHz image with regions selected to measure the integrated flux density. The green ellipse, cyan polygon, and yellow box represent the disk, tail, and extended tail region, respectively.}
    \label{fig:fluxreg}
\end{figure}

To assess the radio spectrum of HCG~97b, we re-imaged LOFAR and VLA data with the same lower UV-cut (250$\lambda$), Briggs weighting ($-0.7$), and outer UV-taper ($10\arcsec$). The resulting images were also convolved to the same resolution ($20\arcsec \times 20\arcsec$) and re-grided to have the same pixel scale (see Figure~\ref{fig:radio_smo20}). The measurements of rms ($\sigma'$) from these images are presented in Table~\ref{tab:radio-data-prop}.

To measure the integrated flux density of radio emission within the disk, tail, and extended tail at three frequency bands, we selected three regions, which are presented in Figure~\ref{fig:fluxreg}. These regions were set to cover the entire disk and tail with emissions above 5$\sigma'$ in each LOFAR and VLA image, and the flux densities from LOFAR and VLA images were measured in each region. The extended tail is not detected at two VLA frequencies, but at nearly the position of the extended tail, we notice a bright point source at 4.86~GHz and far fainter at 144~MHz \footnote{The integrated flux densities of this source (2CXO J234727.3-021917; from NED) are measured as $< 3.13$~mJy, $1.04 \pm 0.10$~mJy, and $0.97 \pm 0.08$~mJy at 144~MHz, 1.4~GHz, and 4.86~GHz, respectively. The derived spectral index is $-0.11$ after fitting the three flux densities with a power-law model, suggesting it could be a blazar with a flat spectrum within the MHz to GHz frequency range \citep{healey07,massaro14,dantonio19} or a variable source on a timescale of fewer than the interval between two VLA observations, i.e., five months.}.
The error on the flux density ($\sigma_{S_{\nu}}$) was estimated taking into account the image noise and calibration uncertainty as follows,
\begin{align}
    \sigma_{S_{\nu}} = \sqrt{(N_{\rm beam} \times \sigma_{\rm rms}^{2}) + (f \times S_{\nu})^{2}} 
    \label{eq_err}
\end{align}
where $N_{\rm beam}$ is the number of beams covering the entire region of interest, $\sigma_{\rm rms}$ is the local rms noise of the image, and $f$ is the flux scale uncertainty. Here, we adopted $f = 10\%$ for LOFAR observation \citep{shimwell22} and $f = 5\%$ for VLA observations \citep{PB17}.

\begin{figure}
    \centering
    \includegraphics[scale=0.6]{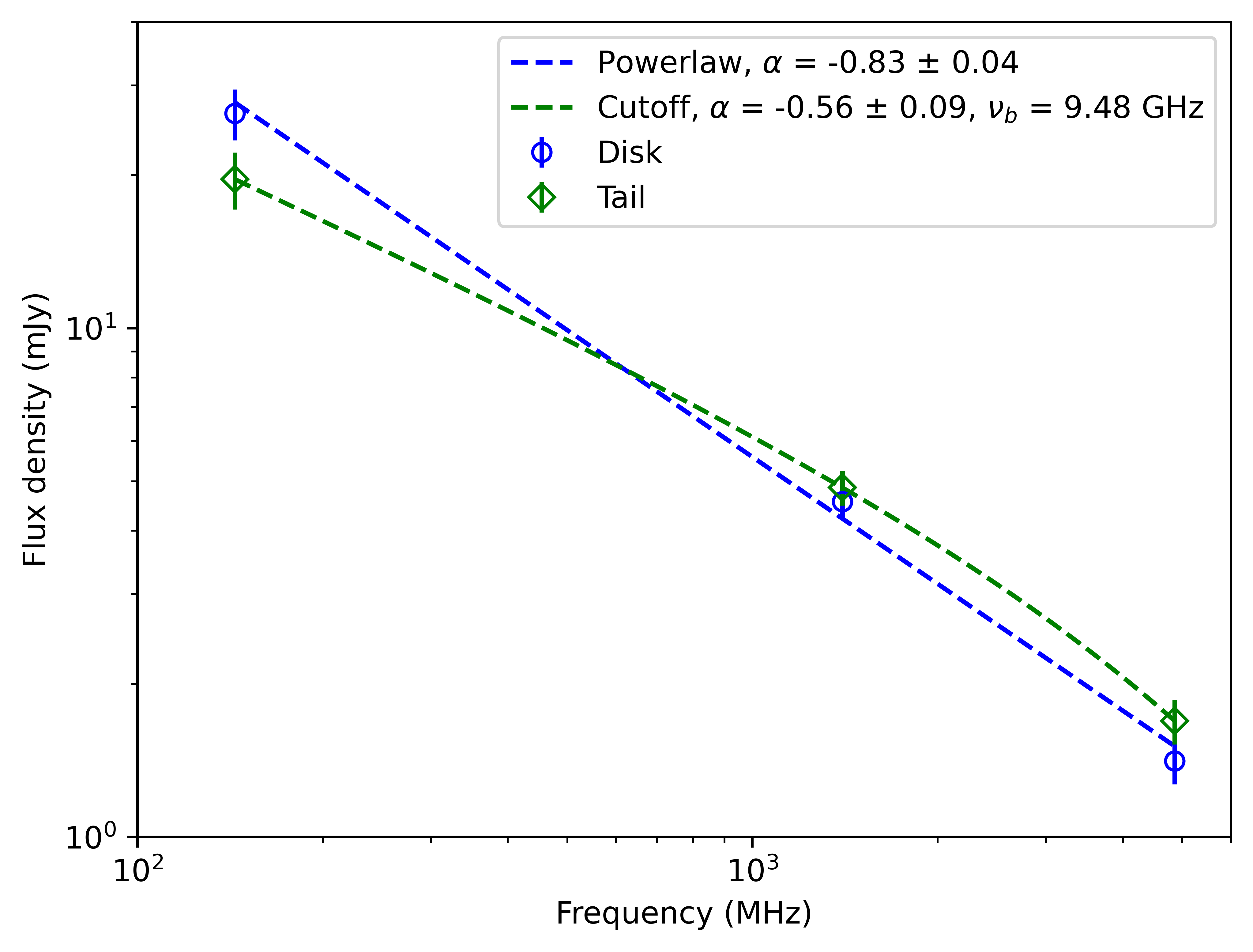}
    \caption{Radio spectra in the disk and tail of HCG~97b at three frequencies. The best fits for the disk and tail are also presented. }
    \label{fig:radio_fit}
\end{figure}

\begin{figure*}
    \centering
    \includegraphics[width=0.94\textwidth]{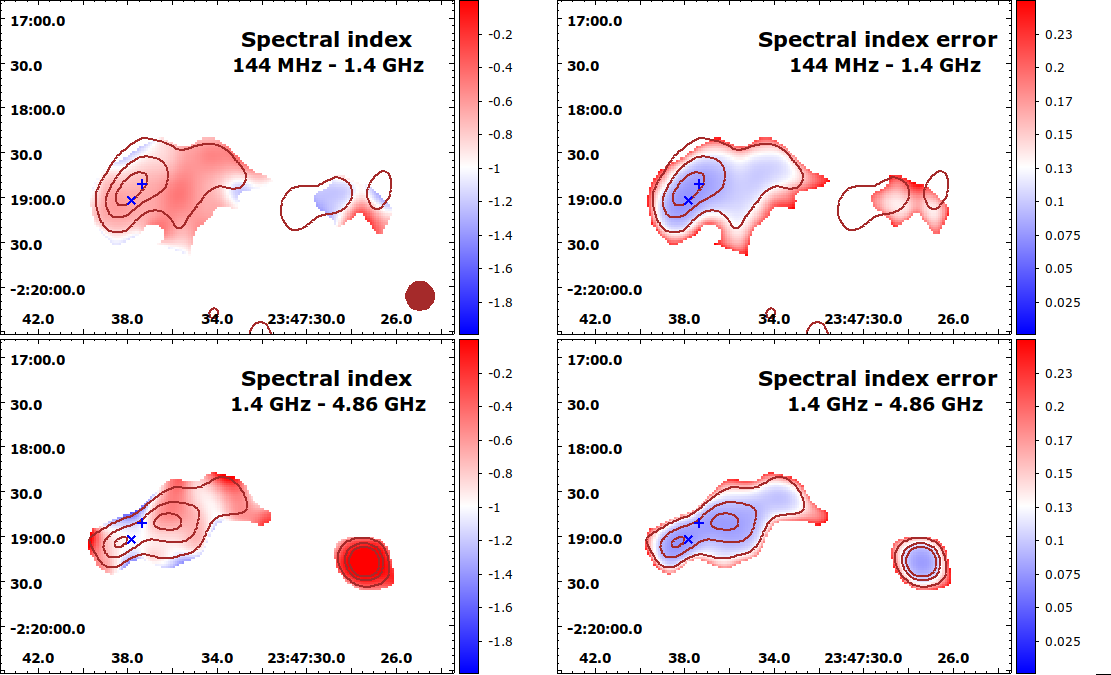}
    \caption{Spectral index (left) and error (right ) maps of radio emission of HCG~97b. LOFAR 144~MHz contours and VLA 1.4 GHz contours derived from the smoothed LOFAR image and VLA image are overlaid on 144 MHz - 1.4 GHz and 1.4 GHz - 4.86 GHz plots, respectively. All contours are starting from $5\sigma'$. The beam shape with a size of $20\arcsec \times 20\arcsec$ is also marked at the right-bottom corner with a filled circle. }
    \label{fig:spcix}
\end{figure*}

In Table~\ref{tab:radio} and Figure~\ref{fig:radio_fit}, we present the flux density measurements and the spectra of the radio emission in the disk and tail of HCG~97b, respectively. The radio spectra in the disk can be well described by a simple power-law model, revealing a spectral index of $-0.83 \pm 0.04$. 
While the radio spectra in the tail exhibit a curvature, which is better characterized by an exponential cut-off model described by the equation:
\begin{align}
    S(\nu) \propto \nu^{\alpha} e^{-\frac{\nu}{\nu_{\rm b}}}
    \label{eq_expcutoff}
\end{align}
where $\alpha$ is the injection spectrum, and $\nu_{\rm b}$ is the break frequency.
The best-fit parameters indicate a relatively flat injection spectrum ($\alpha_{\rm tail} = -0.56 \pm 0.09$) within the tail and a break frequency at $\sim 9.48$~GHz.
The flattening trend in the spectra for both disk and tail is also revealed by the spectral index maps (see Figure~\ref{fig:spcix}). We used smoothed images at 144~MHz, 1.4~GHz, and 4.86~GHz to generate the spectral index maps of radio emission from the disk to tail in two frequency bands, i.e., 144~MHz$-$1.4~GHz and 1.4~GHz$-$4.86~GHz. In each image, only radio emission above $3\sigma'$ was used to calculate the spectral index. Due to the $20\arcsec \times 20\arcsec$ resolution and the galaxy being nearly edge-on, the spectral index distribution within the disk cannot be well-resolved. However, general consistency exists between the spectra ($ -0.8 \lesssim \alpha \lesssim -0.4$) in the low-frequency band and the spectra ($ -1 \lesssim \alpha \lesssim -0.4$) in the high-frequency band. The spectra along the tail are quite uniform and flat ($\alpha \gtrsim -1$) in both frequency bands, and also comparable to those in the disk. The flatness in the tail is potentially influenced by the bright radio blob observed in the tail at 4.86 GHz. We will discuss it in detail in Subsection~\ref{sec:rps-spcix}.

\begin{table}
 \caption{Properties of radio continuum emission of HCG~97b.}
 \label{tab:radio}
 \renewcommand{\arraystretch}{1.3}
 \begin{adjustbox}{max width=0.5\textwidth, center}
 \begin{threeparttable}[h]
 \centering
 \begin{tabular}[c]{lccccc}
  \hline
  Region & $S_{\rm 144~MHz}$ & $S_{\rm 1.4~GHz}$ & $S_{\rm 4.86~GHz}$ & $\alpha_{\rm 144~MHz}^{\rm 1.4~GHz}$  & $\alpha_{\rm 1.4~GHz}^{\rm 4.86~GHz}$  \\
     & (mJy) & (mJy)  &  (mJy) &  &   \\
  \hline
  Disk &  $26.45 \pm 3.02$  &  $4.56 \pm 0.35$  & $1.41 \pm 0.14$  & $-0.77 \pm 0.06$ & $-0.94 \pm 0.05$ \\
  \hline
  Tail  & $19.63 \pm 2.52$  &   $4.86 \pm 0.38$  & $1.69 \pm 0.17$ & $-0.61 \pm 0.07$ & $-0.85 \pm 0.06$  \\
  \hline
  Extended tail  &  $18.09 \pm 2.47$  & $ - $ &  $ - $ & $ - $  & $ - $ \\
  \hline
 \end{tabular}
 \end{threeparttable} 
 \end{adjustbox}
\end{table}

\subsection{CO emission in HCG~97b}
\label{sect:CO}

\subsubsection{CO distribution}
\label{sect:CO-distr}

\begin{figure*}
    \centering
	\includegraphics[width=0.45\textwidth]{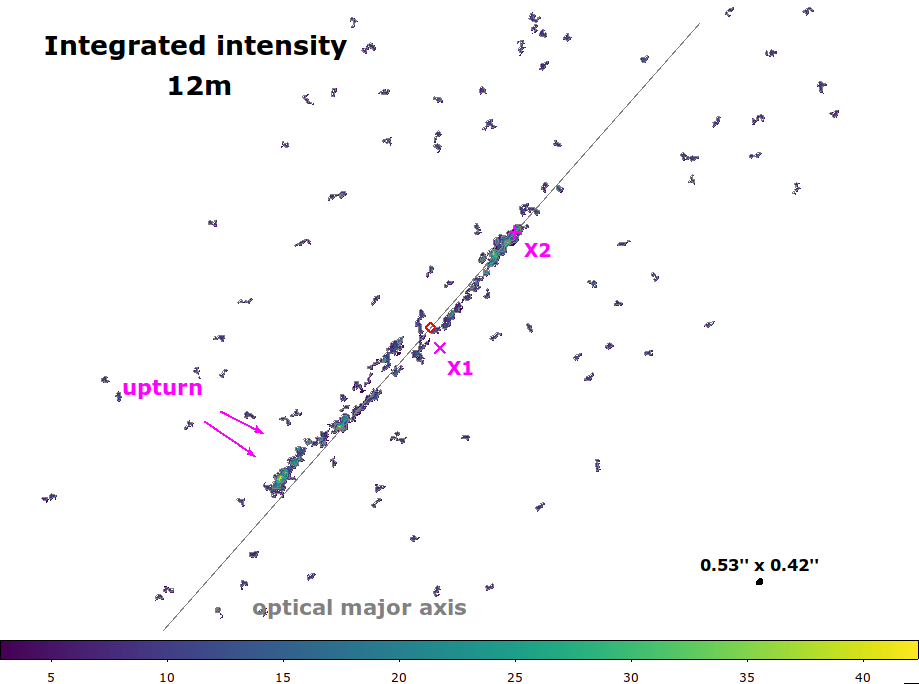}
    \includegraphics[width=0.45\textwidth]{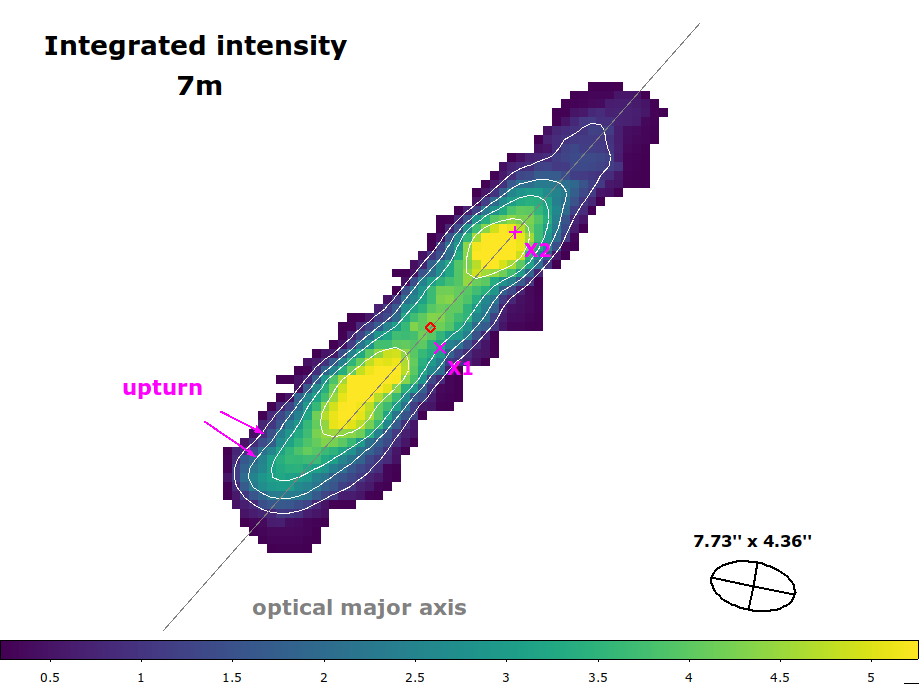}
	\includegraphics[width=0.45\textwidth]{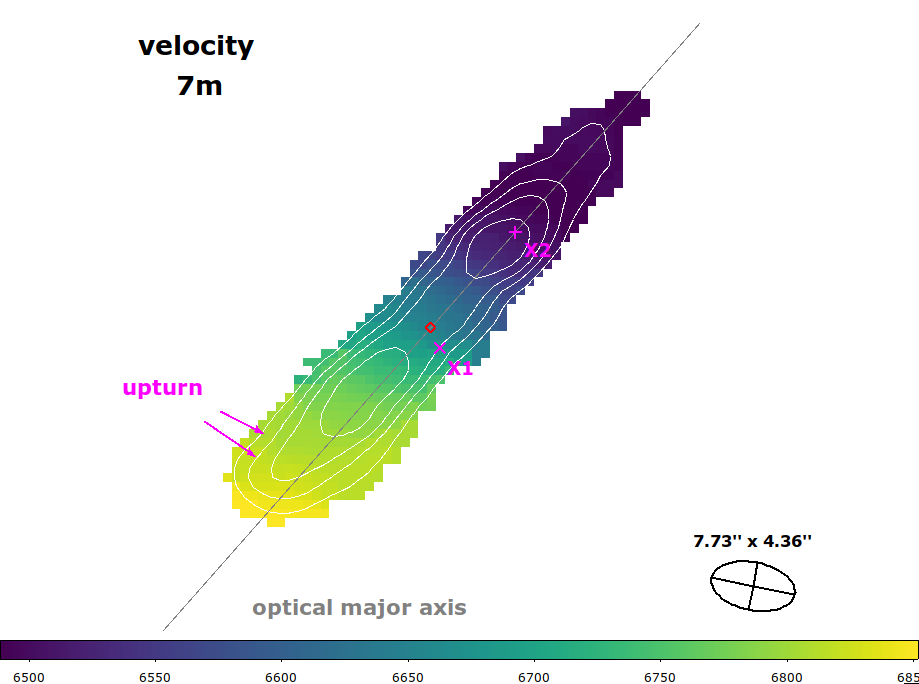}
	\includegraphics[width=0.45\textwidth]{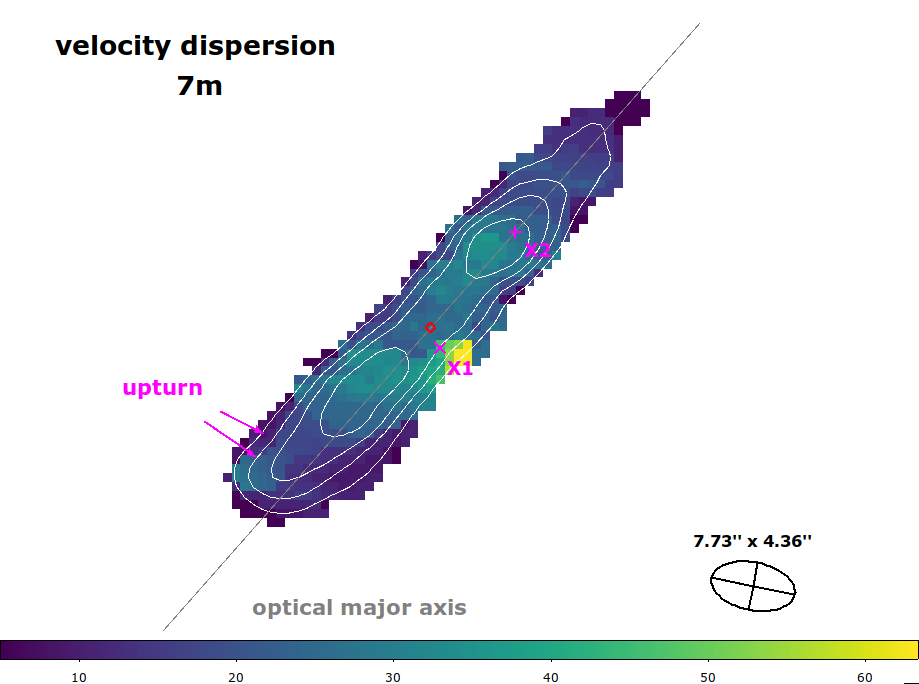}
    \caption{Top: The integrated intensity (unit: $\rm K~km~s^{-1}$) map of CO (2-1) line emission observed with 12m ALMA (left) and 7m ACA (right). The beam sizes for 12m and 7m data are $0.53" \times 0.42"$ and $7.73" \times 4.36"$, respectively. Contours are presented with values of (5, 15, 30) $\rm K~km~s^{-1}$ for 12m data and (1, 2, 3, 4) $\rm K~km~s^{-1}$ for 7m data. Bottom: The intensity-weighted velocity (unit: $\rm km~s^{-1}$) and the corresponding velocity dispersion (unit: $\rm km~s^{-1}$) maps of molecular gas observed with 7m ACA. The galactic centre and two X-ray sources are marked. The major axis of the optical disk is also overlaid.}
    \label{fig:co}
\end{figure*}

The integrated intensity maps of the CO (2-1) emission observed with 12m ALMA and 7m ACA are presented in Figure~\ref{fig:co}. These maps were generated using a signal masking tool \texttt{maskmoment}\footnote{\url{https://github.com/tonywong94/maskmoment}}. 
The mask, within which moment maps are generated, is defined based on a contour with a high significance level of $5\sigma$. This contour is then expanded to include a surrounding lower-significance contour of 3$\sigma$, and is required to cover at least two channels at all pixels.
The CO emission only appears in the disk region and is not detected in the radio tail region. 
In both the 12m and 7m integrated CO intensity maps, the CO emission displays asymmetric morphology. Specifically, the CO (2-1) emission on the southeastern side of the disk is more intense compared to its counterpart, with a slight extension $\sim 1.5\arcsec$ ($\sim 0.67 $ kpc) from the optical major axis toward the northeast (we labelled it as an `upturn' feature in Figure~\ref{fig:co}). 

\begin{figure*}
    \centering
	\includegraphics[width=0.45\textwidth]{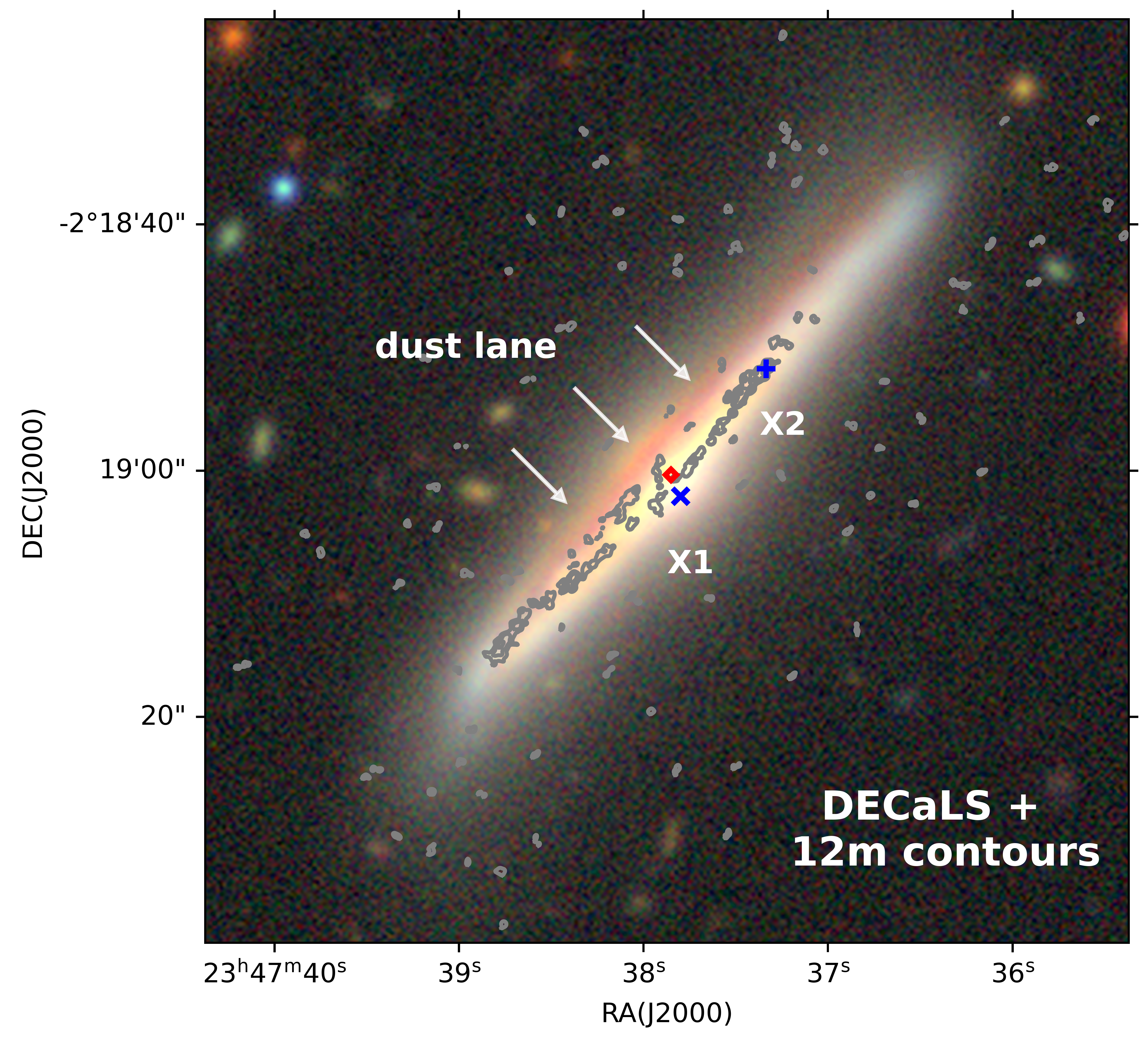}
    \includegraphics[width=0.45\textwidth]{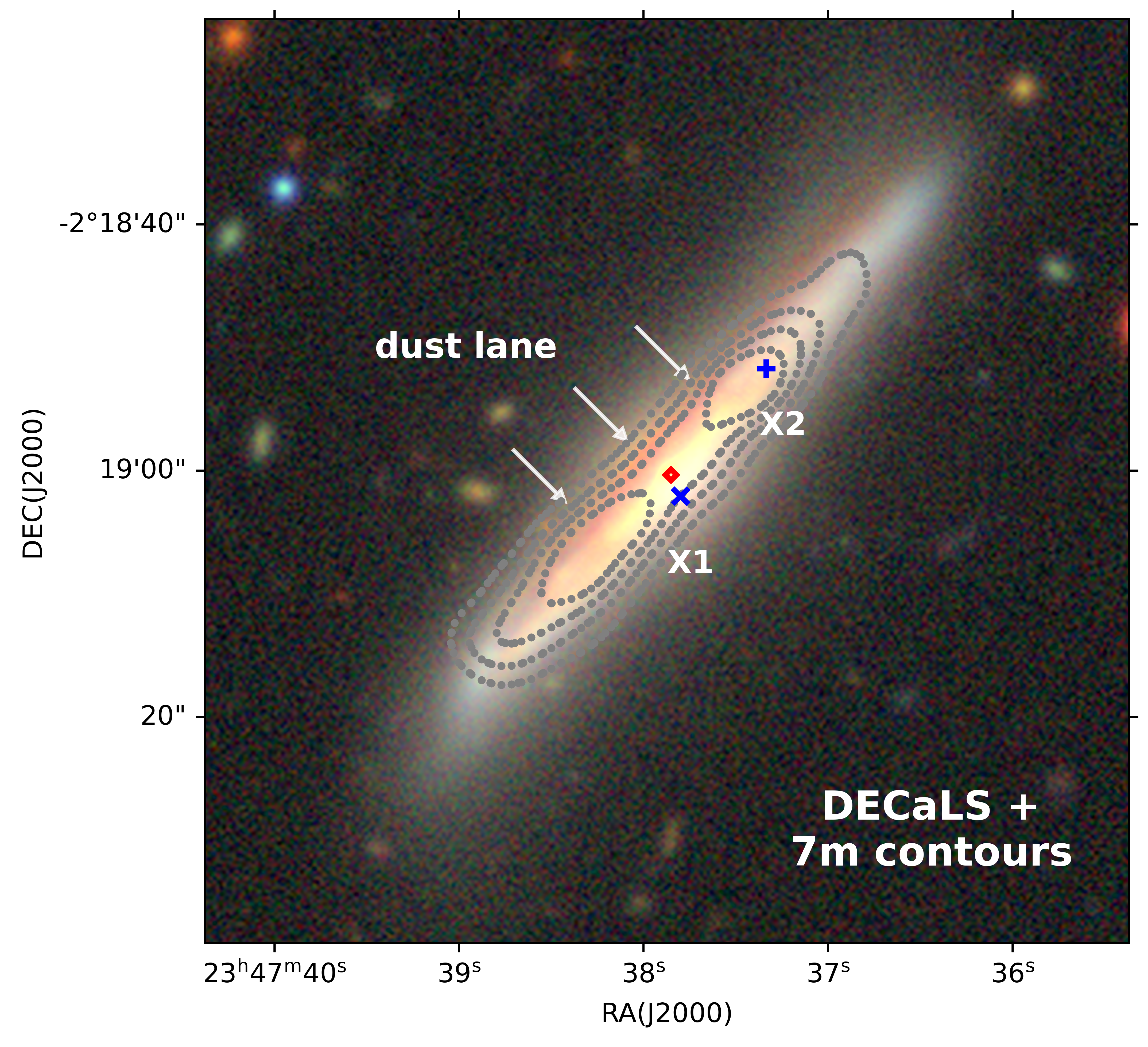}
    \caption{The 12m (left) and 7m (right) integrated CO (2-1) intensity contours are overlaid on the composite DECaLS image. }
    \label{fig:co-decals}
\end{figure*}

To further understand the CO distribution and its relationship with the stellar component, we overlaid the 12m and 7m integrated CO intensity contours on the composite DECaLS image, as shown in Figure~\ref{fig:co-decals}. The optical image with high-resolution 12m ALMA contours reveals that the central CO emission exhibits an elliptical molecular gas distribution, consistent with the minor inclination of the stellar disk. Notably, on the southeast side of the disk, the distribution of the molecular gas in the `upturn' feature is closely aligned with both the dust and the blue stellar components, which show a gentle bent toward the southwest. However, the optical bent is absent on the northwest side of the stellar disk.

\subsubsection{CO kinematics}
\label{sect:CO-kitics}

\begin{figure*}
    \centering
	\includegraphics[width=0.9\textwidth]{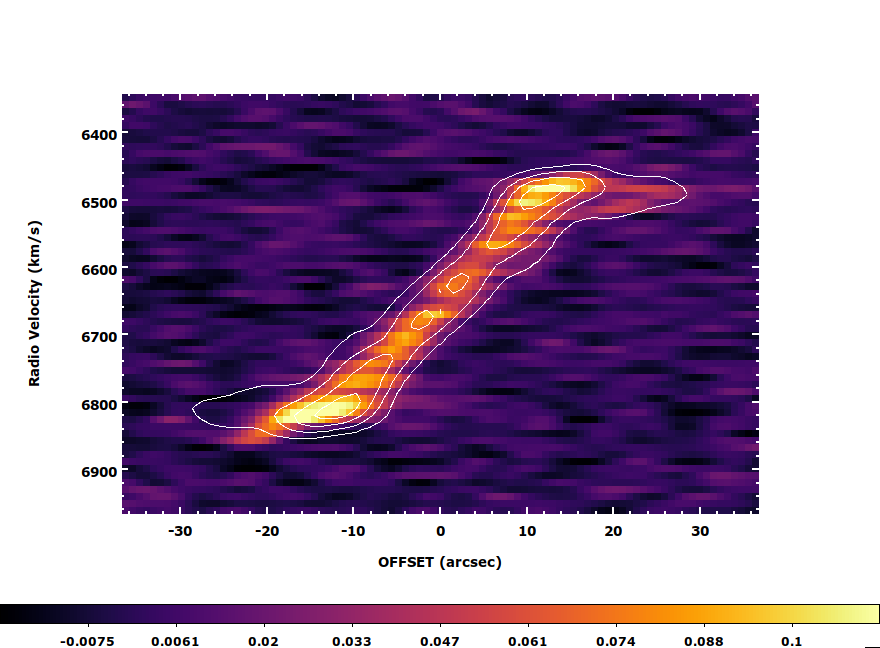}
    \caption{The position–velocity
diagram (PVD) of 7m CO (2-1) emission along the major axis of HCG~97b. Contour levels are 0.02, 0.04, 0.06, and 0.08 $\rm Jy~beam^{-1}$. In contours, a mirror of the right (northwest) side of the galaxy is overlaid on the left (southeast) side.}
    \label{fig:pvd}
\end{figure*}

The intensity-weighted velocity and dispersion maps derived from the 7m data are also presented in Figure~\ref{fig:co}. These maps reveal a maximum and the minimum velocity for the molecular gas of $6877.8~\rm km~s^{-1}$ and $6440.9~\rm km~s^{-1}$, respectively, yielding an overall velocity differential of approximately $437~\rm km~s^{-1}$. This suggests that molecular gas follows the galactic disk rotation at an approximate speed of 218~$\rm km~s^{-1}$.
Furthermore, the Position–Velocity Diagram (PVD) along the major axis is presented in Figure~\ref{fig:pvd}. The PVD of HCG~97b shows asymmetric kinematics; while the northwest (right) side appears undisturbed, the southeast (left) side exhibits anomalous behaviour. Specifically, a sudden velocity spike occurs at $\rm r = 18''$ ($\sim 8$~kpc) following a plateau in the rotation curve between $\rm r = 10''-18''$. To highlight this asymmetry, we overlay the contours from the right side of the galaxy onto the left side, suggesting the presence of a recent perturbation in the outer southeast region of the disk. This pattern resembles the observation in the ram-pressure stripped galaxy NGC~4402 \citep{cramer20}.

\subsection{Molecular Gas Mass Estimation}
\label{sec:SFR}

The total molecular gas mass can be estimated by using $M_{\rm H_{2}} = \alpha_{\rm CO} L_{\rm CO}'$, and the CO line luminosity can be calculated following the equation \citep{SV05}:
\begin{align}
  L_{\rm CO}' = 3.25 \times 10^7~ S_{\rm CO} \Delta \upsilon \nu_{\rm obs}^{-2} D_{\rm L}^{2}~(1+z)^{-3}
\end{align}
where $S_{\rm CO} \Delta \upsilon$ is the CO velocity integrated line flux in $\rm Jy~km~s^{-1}$, $\nu_{\rm obs}$ is the observing frequency in GHz, and $D_{\rm L}$ is the luminosity distance in Mpc. 
Here, we adopted an average CO (2–1)/CO (1–0) line ratio (R21) of 0.64 \citep{brok21} and a CO(1–0)-to-$\rm H_{2}$ conversion factor of $\alpha_{\rm CO} = 4.4$~$\rm M_{\sun}~pc^{-2}~(K~km~s^{-1})^{-1}$ after correcting the contribution of helium with a factor of 1.36 \citep{brok23}. Therefore, the derived CO (2-1) luminosity of HCG~97b is about $3.6 \times 10^8$~$\rm K~km~s^{-1}~pc^{2}$ and the total molecular gas mass $M_{\rm H_{2}}$ is $\sim 2.47\times 10^9$~$\rm M_{\sun}$. 

To exhaust all the molecular gas in HCG~97b at the star formation rate (SFR) of about 1.19~$\rm M_{\sun}~yr^{-1}$ estimated by using the far-infrared (FIR) luminosity ($L_{\rm FIR} =3.39\times 10^{9}$~$\rm L_{\sun}$; obtained from \citealt{martinez12}), the gas depletion time $\tau_{\rm dep}$ is estimated to be around $\sim 2.1\times 10^9$~yr. This is consistent with the average molecular gas depletion time for the disks of normal spiral galaxies \citep[$\sim 2 \times 10^9$~yr; e.g.,][]{bigiel08,leroy13}, suggesting that the current star formation behaviour in HGC~97b is relatively normal.

\section{Discussion}
\label{sec:sect4}

We report the detection of extended radio emission from the spiral galaxy HCG~97b. The LOFAR and VLA continuum images show elongated emission along the plane of the disk and extended toward the northwest with a $\sim 27$~kpc radio tail. The radio tail extends to $\sim 60$~kpc in the LOFAR 144~MHz image, and the outer part of it, i.e., the extended tail (marked in Figure~\ref{fig:radio}), is potentially associated with the X-ray plume from the group centre (or HCG~97a).
The two off-nuclear X-ray sources detected in the spiral galaxy with $2-10$~keV luminosity $\sim 10^{39} - 10^{40}$~$\rm erg~s^{-1}$ have properties that are consistent with ULXs.
The source X2 is mildly obscured ($N_{\rm H}= 2.28 \pm 0.57 \times 10^{22}$~$\rm cm^{-2}$), which is consistent with its location within the northwestern CO emission peak.
Moreover, the CO emission in the galaxy displays asymmetric morphology, with the southeastern side showing greater intensity and alignment with dust and blue stellar components. The PVD reveals asymmetric kinematics, with a sudden velocity spike in the southeast, likely indicative of a recent disturbance such as ram-pressure stripping \citep[e.g.,][]{lee17,cramer20,roberts22b,sardaneta22,cramer23}.

\subsection{Mass estimation of the black holes}
\label{sec:bh}

ULXs are generally considered the most likely candidates for IMBHs. 
The black hole mass could be estimated based on the assumption of the Eddington ratio ($\lambda_{\rm Edd} = L_{\rm bol} / L_{\rm Edd}$), and is given by 
\begin{align}
    M_{\rm BH} & \sim \frac{\kappa_{\rm bol}L_{\rm X}\sigma_{\rm T}}{4\pi G \lambda_{\rm Edd}m_{\rm p} c}\,\notag\\
    &\sim 7.96 \times 10^{4} \left(\frac{\kappa_{\rm bol}}{10}\right) \left(\frac{L_{\rm X}}{1.0 \times 10^{40}\,{\rm erg\,s^{-1}}} \right)\left(\frac{\lambda_{\rm Edd}}{10^{-2}}\right)^{-1}\,{\rm M_{\odot}\,}
    \label{eq_BHmass}
\end{align}
where $\kappa_{\rm bol}$ is the bolometric correction and $L_{\rm X}$ is the $2-10$~keV X-ray luminosity. Here, we scaled $\kappa_{\rm bol}$, $L_{\rm X}$, and $\lambda_{\rm Edd}$ to their typical values. Below, we consider two basic modes of accretion -- radiatively inefficient with $\lambda_{\rm Edd}\leq 10^{-2}$ and the standard-disk accretion with $\lambda_{\rm Edd}\sim 1$. These two accretion modes then determine different ranges for the bolometric-correction factor as a function of $L_{\rm X}$.  

Assuming a radiatively inefficient mode, we used the luminosity-dependent bolometric correction given by $\kappa_{\rm bol} \approx 13 \times (L_{\rm X}/10^{41}~\rm erg~s^{-1})^{-0.37}$, which is derived for low-luminosity advection-dominated accretion flow (ADAF) mode of accretion for LLAGNs \citep{nemmen14}. For the sources X1 and X2, the resulting X-ray bolometric correction factors ($\kappa_{\rm bol}$) are 43.7 and 24.5, and the corresponding bolometric luminosities ($L_{\rm bol}$) are $1.65\times10^{41}$~$\rm erg s^{-1}$ and $4.41\times10^{41}$~$\rm erg s^{-1}$, respectively. 
For the ADAF model of accretion, the Eddington ratio is required to be small enough $\lambda_{\rm Edd} \leq 10^{-2}$. If we adopt the upper limit of the Eddington ratio $\lambda_{\rm Edd}$ ($\sim 10^{-2}$), the lower limits of black hole masses for sources X1 and X2 can be obtained as $1.3\times10^5$~$\rm M_{\sun}$ and $3.5\times10^5$~$\rm M_{\sun}$, respectively. In the radiatively inefficient mode, the Eddington ratio is not expected to be smaller than $\lambda_{\rm Edd}\leq 3.5 \times 10^{-3}$ since then the X2 mass exceeds $10^{6}\,M_{\odot}$, which is inconsistent with the absence of the major merger fingerprints.
If the accretion proceeds effectively via a standard disk with an Eddington ratio close to the unity ($\lambda_{\rm Edd} \sim 1$), the X1 and X2 masses would be pushed even lower. In this case, we used the bolometric correction $\kappa_{\rm bol} = 7\times (L_{\rm X} / 10^{42}~\rm erg s^{-1})^{0.3}$ for a thin disk following \citet{netzer19} and derived the $\kappa_{\rm bol}$ as 1.3 and 2.1 for the X1 and X2 $2-10$~keV luminosity, respectively. Therefore, the black hole masses are 39.5~$\rm M_{\sun}$ and 300.4~$\rm M_{\sun}$ for sources X1 and X2, respectively. This shows that the X2 is consistent with being an IMBH, while X1 could even be a stellar-mass black hole accreting close to or even exceeding an Eddington rate. While if a super-Eddington accretion ($\lambda_{\rm Edd} \approx 10^{2}$) is assumed, the black hole mass of X1 and X2 are 0.4~$\rm M_{\sun}$ and 3.0~$\rm M_{\sun}$, respectively, placing them both within the stellar-mass black hole range. The measurements of black hole masses are also summarised in Table~\ref{tab:bh_mass}.

A further way to ascertain the black hole mass is to use the Fundamental Plane correlation between X-ray luminosity and radio luminosity \citep{merloni03,plotkin12,dong14,gultekin19}. Since we detected no compact radio sources spatially co-aligned with X1 and X2, we estimated limits on the integrated radio luminosities using the same regions as in the X-ray spectral analysis (see Subsection~\ref{sect:xray}). The upper limits of the radio luminosities for X1 and X2 are $L_{\rm X1,~4.8~GHz} =4.03\times 10^{36}$~$\rm erg~s^{-1}$ and $L_{\rm X2,~4.8~GHz} =3.16\times 10^{36}$~$\rm erg~s^{-1}$.
Following the Fundamental Plane correlations proposed by \citet{gultekin19}, $\rm log({\it M_{\rm BH}}/10^8 ~\rm M_{\sun}) = (0.55\pm0.22)+(1.09\pm0.10)log({\it L_{\rm R,~ 5.0~GHz}}/10^{38}~\rm erg~s^{-1}) + (-0.59_{-0.15}^{+0.16})log({\it L_{\rm X, ~2-10~keV}}/10^{40}~\rm erg~s^{-1})$, the black hole masses are constrained as $\leq (1.9 \pm 1.3)\times 10^{7}$~$\rm M_{\sun}$ for X1 and $\leq (5.8\pm 3.9)\times 10^6$~$\rm M_{\sun}$ for X2, which also suggests the X2 could fall into the IMBH mass range.

\begin{table*}
 \centering
 \caption{Estimations of black hole masses for two ULXs, X1 and X2, for different Eddington ratios and the fundamental plane relation. In the last row, we distinguish different accretion modes for the corresponding Eddington ratios.}
 \label{tab:bh_mass}
 \renewcommand{\arraystretch}{1.5}
 \begin{tabular}[c]{ccccc}
  \hline
  X-ray source  &  \multicolumn{3}{c}{ $M_{\rm BH}$ ($\rm M_{\sun}$)}  \\
  \cline{2-4}
     & $\lambda_{\rm Edd} = 10^2$  & $\lambda_{\rm Edd} = 1$ & $\lambda_{\rm Edd}=10^{-2}$ &  Fundamental Plane  \\
  \hline
  X1    &  $0.4$  & $39.5$ & $1.3\times 10^5$    & $\leq (1.9 \pm 1.3)\times 10^{7}$ \\
  \hline
  X2   &  3.0   & 300.4 & $3.5\times 10^5$   & $\leq (5.8\pm3.9)\times 10^6$ \\  
  \hline
  \hline
  accretion mode & slim disk & slim/standard disk &  ADAF &   \\
  \hline
 \end{tabular}
 \end{table*}

\subsection{Ram-pressure stripped radio tail}
\label{sec:RPS}

The observed extended, one-sided asymmetric shape of the radio tail, in conjunction with the asymmetry in CO morphology and kinematics, suggests that the galaxy may be experiencing ram pressure, likely giving rise to the formation of a stripped radio tail.

\subsubsection{Ram pressure direction}
\label{sec:RPS-direction}

\begin{figure}
    \centering
	\includegraphics[scale=0.3]{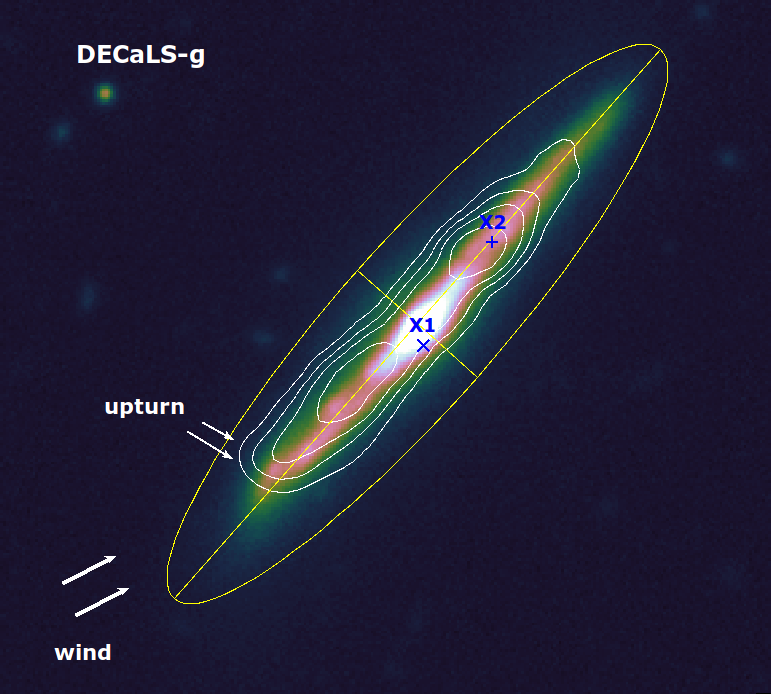}
	\includegraphics[scale=0.3]{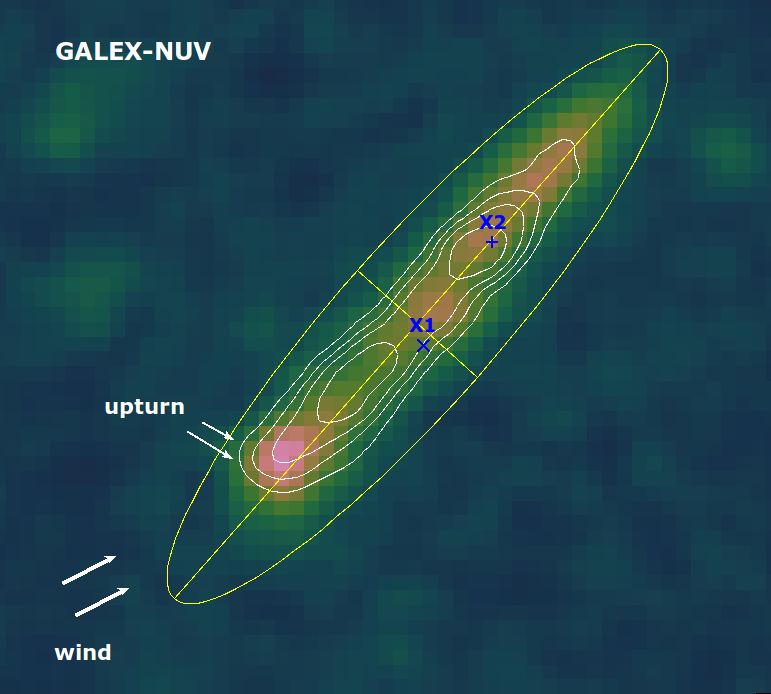}
    \caption{DeCaLS g-band image (top) and GALEX NUV image (bottom) with ALMA integrated intensity contours. The feature `upturn' and two X-ray sources are marked. The IGrM wind due to the ram pressure is indicated. The major and minor axes of the optical disk are also overlaid. }
    \label{fig:upturn}
\end{figure}

The stellar disk and molecular gas distributions in HCG~97b (see Figure~\ref{fig:co-decals} and Figure~\ref{fig:upturn}) exhibit a bent or `upturn' feature which offsets to the north of the major axis. Correspondingly, in the Galaxy Evolution Explorer (GALEX) near-UV (NUV) image (the bottom panel of Figure~\ref{fig:upturn}), the peak NUV emission coincides with this `upturn', implying a similar offset distribution for young stars. Given that HCG~97b is not a real edge-on galaxy --- with an estimated inclination angle of $85\degree - 88\degree$, this `upturn' feature is likely not an extraplanar feature but rather indicative of a non-axisymmetric disk feature. These observed features in HCG~97b resemble those in the Virgo cluster galaxy NGC~4654. NGC~4654 has two asymmetric spiral arms, where the shorter arm ends in a bright star-forming region at the outer edge of the gas disk, aligning with the leading side of the ram pressure interaction \citep{KK04,chung09}. The distributions of stars and gas in NGC 4654 are believed to be the result of combined effects, including gravitational interaction ($\sim$ 500 Myr ago) and ram pressure stripping (ongoing; \citealt{vollmer03}). Therefore, HCG~97b could resemble the NGC~4654 but viewed from a different angle.  
Moreover, the radio tail of HCG~97b is extended toward the northwest and radio contours on the southeast side are compressed and tight compared with those on the other side (see Figure~\ref{fig:radio}), suggesting the direction of the ram pressure should be from the southeast (see the indication of the ram pressure direction in Figure~\ref{fig:upturn}).

This scenario is also indicated by the elongated X-ray plume from the central galaxy HCG~97a toward the southeast, presented in Figure~\ref{fig:overall} and Figure~\ref{fig:softxray-radiospc}. Since the X-ray plume is roughly extended in the direction of HCG~97b, it is likely an X-ray tail stripped from the hot gas halo of HCG~97a or the central IGrM during the past interaction between HCG~97a and HCG~97b. This is also the case for M86, with a possible encounter of NGC~4438 first reported in \citet{Kenney08}, and NGC~4438 shows a clear sign of tidal interaction. However, the disk morphology of HCG~97b is not strongly perturbed, suggesting that the X-ray plume might have been formed in the past by the interaction between HCG~97a and other galaxies. Alternatively, it could also be a scenario that HCG~97b with a sub-group fell into the main group HCG~97 in the past, and the X-ray halo of the sub-group was grabbed by the main group during the first pericentric passage, forming an X-ray plume pointing to the motion direction. We also notice that the extended radio tail detected in the LOFAR 144~MHz image coincides with the end of the X-ray plume. This scenario could be that the striped tail of HCG~97b encountered the X-ray plume, and then the low-frequency radio emission was "re-ignited" by the turbulence or compression possibly triggered by the bulk motion in the X-ray plume. Considering a typical radiative time range of CRe at 144~MHz ($\sim 1 - 2 \times 10^8$~yr; \citealt{basu15}), the stripped scale of the radio tail could be estimated as $\sim 36-74$~kpc after assuming that the velocity of CRe follows the relative galaxy velocity ($v \approx 360$~$\rm km~s^{-1}$), which is the difference between the velocities of HCG~97b ($cz_{\rm HCG~97b}=6940$~$\rm km~s^{-1}$; \citealt{hickson92}) and the group mean ($cz_{\rm HCG~97}=6579$~$\rm km~s^{-1}$; \citealt{jones23}). It suggests that there could exist old radio plasma at the 60~kpc radio tail still emitting radio emission at 144~MHz.

\subsubsection{Capabilities of ram-pressure stripping and gravitational interaction}
\label{sec:RPS-capability}

Theoretically, ram-pressure stripping and gravitational or tidal interactions are inevitable phenomena for galaxies orbiting within a galaxy group environment. To quantify these effects for HCG~97b, we conduct approximate calculations to estimate the relative capabilities of the ram-pressure stripping and gravitation interaction.

\begin{figure*}
    \centering
	\includegraphics[width=\textwidth]{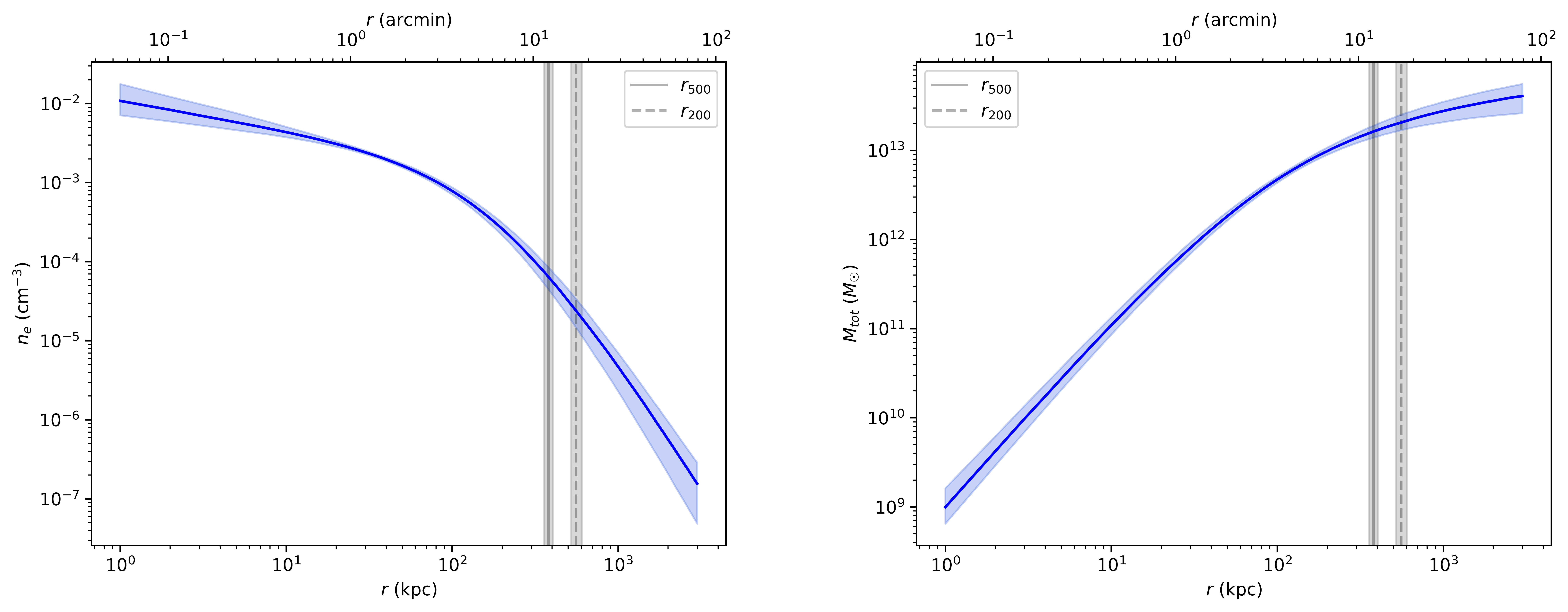}
    \caption{The distributions of the electron number density (left) and total mass (right) of the galaxy group HCG~97 using the RTI model \citep{zhu16,zhu21}.}
    \label{fig:density}
\end{figure*}

To derive the distributions of the ram pressure and radial tidal interaction, we first determine the electron number density and total mass of the HCG~97 group by using a revised thermodynamical ICM (RTI) model \citep[see][for details of the RTI model]{zhu16,zhu21}, and present the results with 68\% confidence level in Figure~\ref{fig:density}.

\begin{figure*}
    \centering
	\includegraphics[width=\textwidth]{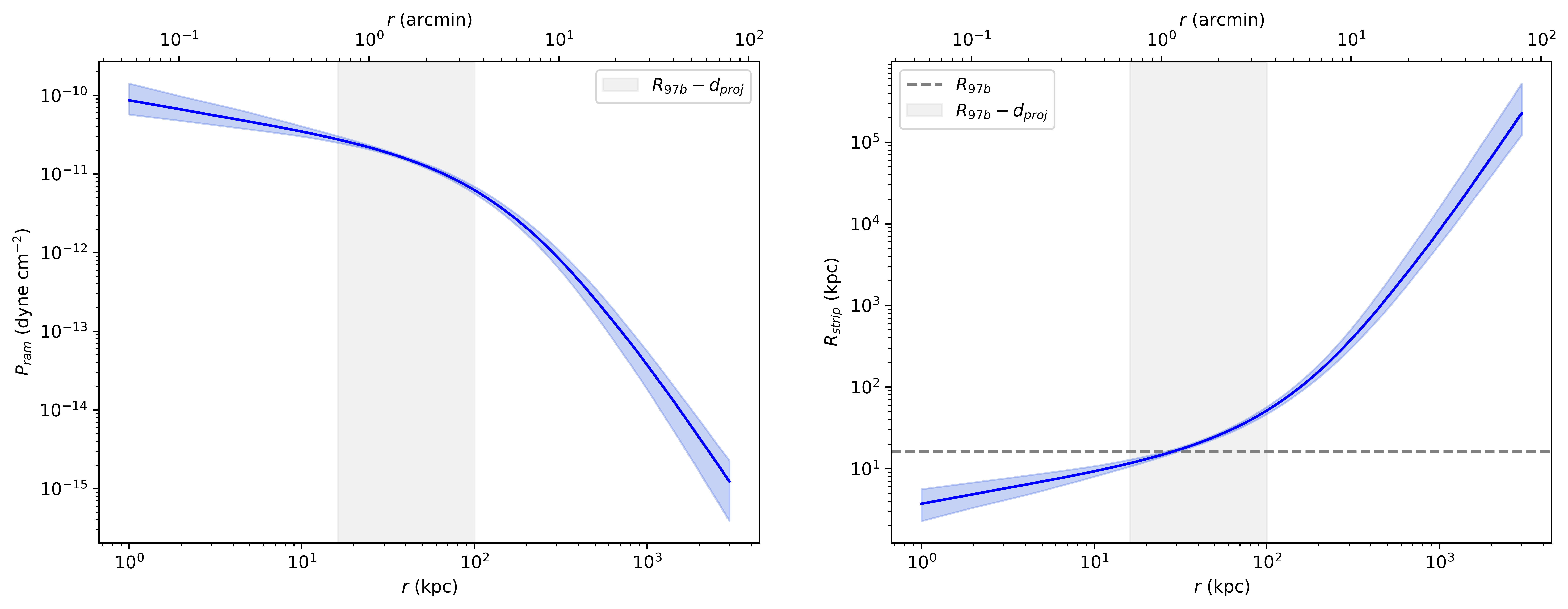}
    \caption{The ram pressure (left) and corresponding stripped radius (right) of HCG~97b as the function of the distance between the group centre and HCG~97b. The shadow region represents the range between the radius of HCG~97b ($\rm R_{97b}$) and the current projected separation between the group centre and HCG~97b ($\rm d_{proj}$). The radius of HCG~97b is also marked with a horizontal dash line. }
    \label{fig:P_ram}
\end{figure*}

Following the method described in \citet{jachym14}, the ram pressure can be estimated as
\begin{align}
  P_{\rm ram} = \rho_{\rm gas, IGrM} \upsilon^{2} =  \mu_{\rm e} m_{\rm u} n_{\rm e, IGrM} \upsilon^{2}
\end{align}
where $\rho_{\rm gas, IGrM} = \mu_{\rm e} m_{\rm u} n_{\rm e, IGrM}$ is the gas mass density of the group at the position of the galaxy, $m_{\rm u}$ is the atomic mass unit, and $\upsilon$ is the three-dimensional (3D) infalling velocity of HCG~97b. Since the relative galaxy velocity ($\sim 360$~$\rm km~s^{-1}$) is lower than the group velocity dispersion ($\sigma_{\rm HCG~97} \approx 371.5$~$\rm km~s^{-1}$; \citealt{hickson92}), we adopt the group velocity dispersion here to approximate the 3D infalling velocity $\upsilon$ ($\approx \sqrt{3} \sigma_{\rm HCG~97}$). 
Then, we compare the ram pressure acting on the galaxy's atomic gas with the gravitational restoring force in the galaxy. The latter can be approximated by the centrifugal force at the galaxy's outer radius ($R$) with an expression of $a=v_{\rm rot}^{2} / R$, where $v_{\rm rot}=218$~$\rm km~s^{-1}$ is the rotation velocity obtained from the ALMA data (see Subsection~\ref{sect:CO}). Assuming a typical column density ($\Sigma_{\rm H\textsc{i}} \sim 10~\rm M_{\sun}~pc^{-2}$) of the galaxy's atomic gas and a flat rotation curve, the radius $R$ can be associated with the stripping radius, which can be estimated as
\begin{align}
  R_{\rm strip} = v_{\rm rot}^{2} \Sigma_{\rm H\textsc{i}}  / P_{\rm ram} 
\end{align}
The derived ram pressure and stripped radius are present in Figure~\ref{fig:P_ram}.

\begin{figure*}
    \centering
	\includegraphics[width=\textwidth]{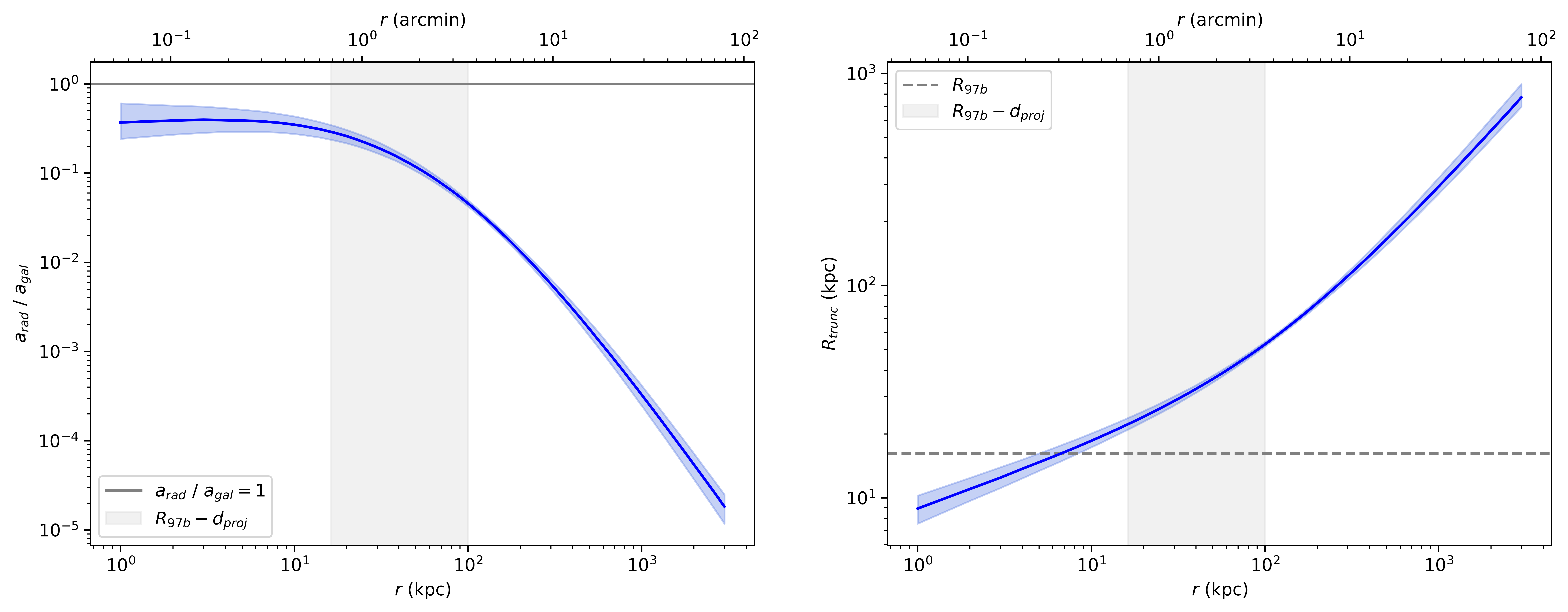}
    \caption{The ratio between the radial tidal acceleration and the internal galaxy acceleration (left) and the truncation radius (right) due to the group gravitational well as the function of the distance between the group centre and HCG~97b. The shadow region represents the range between the radius of HCG~97b ($\rm R_{97b}$) and the current projected separation between the group centre and HCG~97b ($\rm d_{proj}$). The radius of HCG~97b is also marked with a horizontal dash line. }
    \label{fig:a_tidal}
\end{figure*}

Following the calculations provided by \citet{HB96} and \citet{cortese07}, we calculate the strength of the galaxy-group gravitational interaction as HCG~97b falling into the group potential well. It can be quantified based on the comparison between the radial tidal acceleration ($a_{\rm rad}$) and internal galaxy acceleration ($a_{\rm gal}$), which are given by
\begin{align}
  a_{\rm rad} = G M_{\rm pert} \left[ \frac{1}{r^2} - \frac{1}{(R+r)^2} \right] 
\end{align}
\begin{align}
  a_{\rm gal} = \frac{G M_{\rm dyn}}{R^2}  
\end{align}
where $M_{\rm pert}$ is the total mass of the perturber (i.e., galaxy group HCG~97) within $r$, $M_{\rm dyn}$ is the dynamic mass of the perturbed galaxy (i.e., HCG~97b), $R$ is the radius of the perturbed galaxy, and $r$ is the separation between the perturber and perturbed galaxy. The radial tidal acceleration has the effect of pulling the matter at either side of the disk away from the galaxy and approaching the perturber. We estimate the dynamic mass of HCG~97b ($M_{\rm dyn} \sim 7\times 10^{11}~\rm M_{\odot}$) based on the stellar mass - halo mass relationship provided by \citet{BWC13}. If the radial tidal acceleration is stronger than the internal galaxy acceleration, the matter on the infalling galaxy is able to be stripped. The corresponding stripped radius (i.e., truncation radius $R_{\rm trunc}$) can be estimated by 
\begin{align}
  R_{\rm trunc} \approx r \left( \frac{M_{\rm dyn}}{M_{\rm pert}(<r)} \right)^{1/3}  
\end{align}
We present the ratio between two accelerations ($a_{\rm rad}/a_{\rm gal}$) and the truncation radius as a function of the distance $r$ in Figure~\ref{fig:a_tidal}. These results agree well with the case of the galaxy with a stellar mass of $10^{10}~\rm M_{\odot}$ within a group environment (see the green solid lines in their Figure 7 and Figure 8) provided in \citet{BFS22}.

From the plots in Figure~\ref{fig:P_ram} and Figure~\ref{fig:a_tidal}, it can be observed that if HCG~97b is an infalling galaxy approaching the group centre, neither ram pressure nor tidal interaction would have a significant impact on HCG~97b, because both the stripped radius and truncation radius are quite larger than the radius of HCG~97b. Therefore, it confirms that HCG~97b has undergone the first pericentric passage and is currently receding from the group centre. Given that the optical disk of HCG~97b does not exhibit obviously disturbed morphology (only a possible weak warp shape), it is plausible to assume that the separation between HCG~97b and the group centre during the first pericentric passage was not extremely close, perhaps greater than the disk radius of HCG~97b ($R_{\rm 97b} = 1.23' = 16.2$~kpc). It is more probable that HCG~97b has experienced the closest distance of $\gtrsim 20-30$~kpc between the group centre and HCG~97b. At this distance, both hydrodynamic and gravitational processes likely played roles in gas stripping. The gravitational interaction might cause a possible mild warp of the disk and possibly flatten the galaxy's gravitational potential well, enhancing the efficiency of ram pressure. This is supported by the curve in Figure~\ref{fig:a_tidal}, which suggests a mild gravitational interaction effect with values of $a_{\rm rad}/a_{\rm gal} \sim 10^{-1}$ for $r \approx 20-30$~kpc. It is also worth noting that the real ram pressure would be even higher because the current relative velocity of HCG~97b has been largely reduced after the first pericentric passage. 

More recently, \citet{jones23} used VLA data to measure the H\textsc{i} emission of Hickson Compact Groups (HCGs) and reported that for the group of HCG~97, the H\textsc{i} emission is only detected in the member galaxy HCG~97b and its morphology is more extended at the northern side of the disk, implying that HCG~97b is likely a new, gas-rich galaxy infalling into an evolved group. They also measured the $M_{\rm H\textsc{i}}$ deficiency Def($M_{\rm H\textsc{i}}$)~$=0.79$ of HCG~97b, which is the logarithmic difference between the predicted value (log~$M_{\rm H\textsc{i},pred}=9.47$~$\rm M_{\sun}$) and observed one (log~$M_{\rm H\textsc{i}}=8.68$~$\rm M_{\sun}$). \citet{martinez12} also measured Def($M_{\rm H\textsc{i}}$) but inferred a higher value of Def($M_{\rm H\textsc{i}}$)~$=1.53$. The difference between the $M_{\rm H\textsc{i}}$ deficiency in the two papers is probably because \citet{jones23} used the improved imaging (e.g., multi-scale clean) and masking (e.g., \texttt{SoFiA}) methods to recover as much flux of HCG~97b as possible, thus obtaining a smaller deficient value. Nonetheless, both works suggest that a considerable part of H\textsc{i} was removed from HCG~97b due to the possible interaction between the infalling galaxy and IGrM. 
Furthermore, \citet{rasmussen08} modelled ram-pressure stripping in HCG~97b and reported that up to $2 \times 10^9$~$\rm M_{\sun}$ of H\textsc{i} could have been stripped from a galaxy with a stellar mass of $3.2 \times 10^{10}$~$\rm M_{\sun}$. This value is roughly consistent with the observed deficiency of HCG~97b \citep{jones23}.

\subsubsection{Atypical spectral indices in the radio tail}
\label{sec:rps-spcix}

\begin{figure}
    \centering
    \includegraphics[scale=0.6]{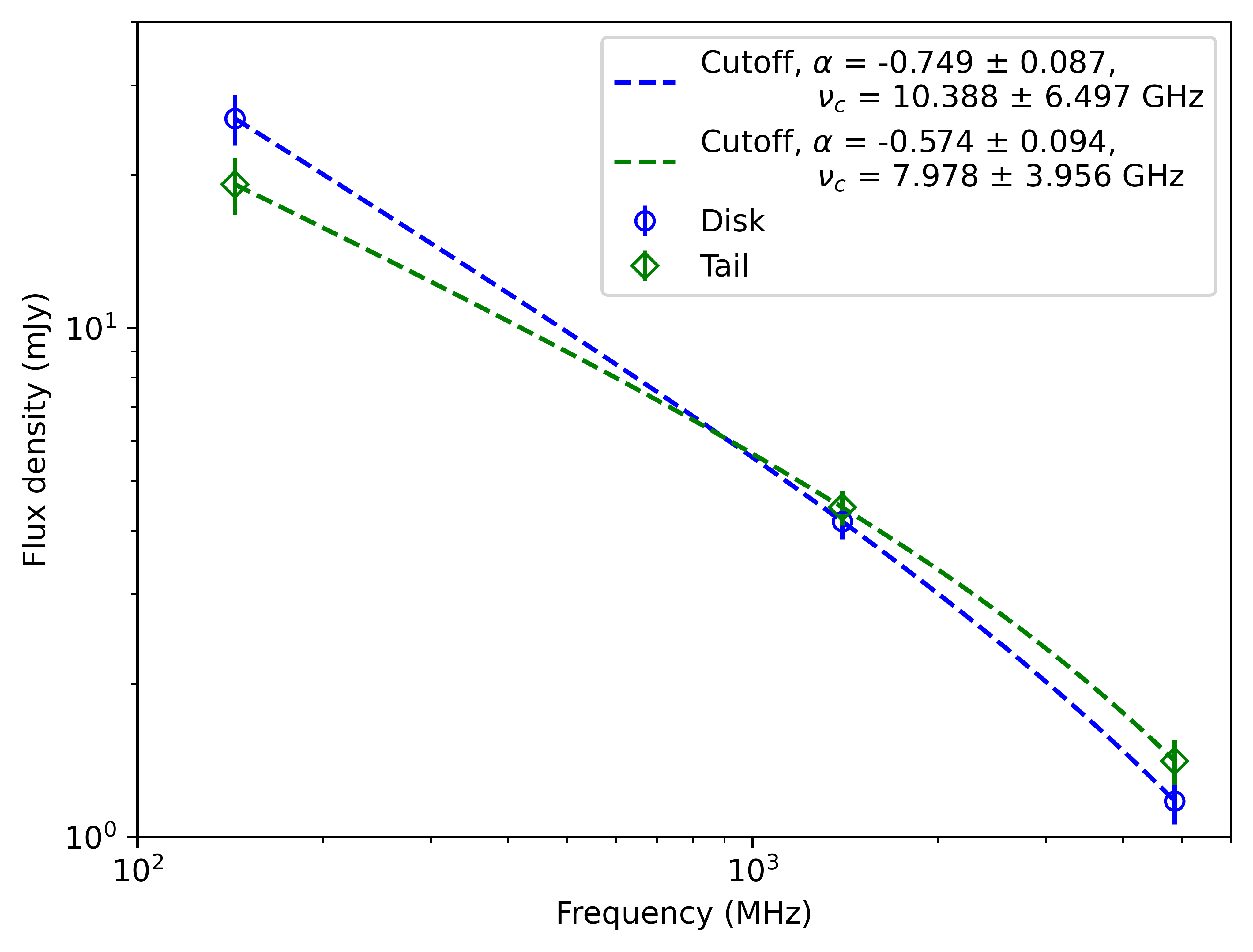}
    \caption{Radio synchrotron spectra in the disk and tail of HCG~97b after the thermal emission correction. The best fits with the exponential cut-off model are also presented. }
    \label{fig:radio_fit2}
\end{figure}

In the case of HCG~97b, radio spectral indices in the radio tail differ from typical ram-pressure stripped tails, showing flatter values comparable to those in the radio disk, both in the integrated spectra and in the spectral index maps. This contradicts the expected steeper spectral indices ($\alpha < -1$) in stripped tails due to synchrotron radiative cooling \citep[i.e.,][]{vollmer04,chen20,muller21,vollmer21,roberts22,ignesti22}. We note that while nonthermal emission (synchrotron emission) dominates the radio emission at $\nu \lesssim 10$~GHz in most nearby galaxies, thermal emission (free-free emission) could also contribute a fraction of $f_{\rm th} \approx 1\% - 40\%$ of the observed total radio emission \citep[][and the references therein]{klein18}. Since there is a lack of H$\alpha$ data to estimate the thermal emission for HCG~97b directly, we approximate the non-thermal emission by assuming thermal emission fractions of 2.3\%, 8.6\% and 16.6\% at 144 MHz, 1.4 GHz and 4.86 GHz, respectively, based on Equation 8 provided by \citet{tabatabaei17}. We present the estimated nonthermal spectra in the disk and tail, along with the best fits with the exponential cut-off model in Figure~\ref{fig:radio_fit2}. The nonthermal spectra in the disk indicate a more curved shape and an injection spectral index of $\alpha_{\rm nth}^{\rm disk} \approx -0.75 \pm 0.09$. While the new spectra in the tail suggest an unchanged injection spectral index ($\alpha_{\rm nth}^{\rm tail} \approx -0.57 \pm 0.09$ ), but a lower break frequency. The spectrum break frequency in the tail ($\nu_{\rm c}^{\rm tail} \approx 8.0 \pm 4.0$~GHz) appears to be smaller than the one in the disk ($\nu_{\rm c}^{\rm disk} \approx 10.4 \pm 6.5$~GHz), suggesting a slight difference in a radiative timescale which can be derived with the expression presented in \citet{miley80}: 
\begin{align}
    t_{r} \propto 3.2 \times 10^{10} \frac{B^{1/2}}{B^2 + B_{\rm CMB}^{2}} \frac{1}{\sqrt{\nu_{\rm c}(1+z)}} ~{\rm yr}
    \label{eq_t_rad}
\end{align}
where $B_{\rm CMB}=3.25(1+z)^2$~$\rm \mu G$ is the equivalent CMB magnetic field and $\nu_{\rm c}$ is the cut-off frequency in units of MHz. Assuming a typical magnetic field of $\sim 10$~$\rm \mu G$ within the disk and $\sim 5$~$\rm \mu G$ within the tail \citep{tabatabaei17,muller21,ignesti22}, the radiative times are estimated at $8.8\pm2.7 \times 10^6$~yr and $2.2\pm0.5 \times 10^7$~yr for the disk and tail, respectively.

The observed flat spectrum in the tail with its longer radiative timescale indicates that the CRe within the tail may have undergone rapid displacement or uplift from the disk, to an extent where the typical spectral aging effects are not observed.
To quantify this displacement, we adopt a derived age of approximately $2.2\pm0.5 \times 10^7$~yr for the CRe in the tail and a projected tail length of approximately 30~kpc. Consequently, we infer an expected velocity for the radio plasma transported from the disk to the tail of roughly 1300~$\rm km~s^{-1}$. This velocity is much higher than the velocity measured in ram-pressure stripped galaxies ($100-600~\rm km ~s^{-1}$; \citealt{ignesti23}), and aligns more closely with scenarios observed in head-tail galaxies where the plasma is accelerated out of the galaxy by the AGN \citep[e.g.,][]{ignesti20,edler22}. 
Given that the injection spectral indices for both disk and tail are consistent with the characteristics of fresh injected plasma typically associated with AGN activity, and taking into account the proximity of the two bright radio blobs detected at 4.86~GHz to the X2 source, we propose that the CRe may have been uplifted through a mechanism involving ULX feedback --- potentially linked to IMBH activity, in addition to the conventional ram-pressure stripping process. In the upcoming section (Subsection~\ref{sec:IMBH-induced}), we will discuss the potential contribution of IMBH activity in more detail. 
Furthermore, the absence of spectral aging effects outside of the disk could also be attributed to efficient adiabatic expansion losses. The energy loss due to adiabatic expansion is linearly proportional to the CRe energy (E), while losses from synchrotron are proportional to the square of the energy ($\rm E^2$; \citealt{longair11}). Considering that the thermal pressure in the galaxy group environments should be lower than in the ICM, the adiabatic losses might be more efficient since the ISM can expand more easily.

\subsection{Enhanced ram-pressure stripping due to an accreting IMBH?}
\label{sec:IMBH-induced}

A peculiarity of the spiral galaxy HCG~97b is the presence of two off-nuclear X-ray sources, with the X2 source being more distant from the galactic nucleus in projection, i.e. $\sim 5.2$~kpc from the galactic nucleus. Given the fact that no clear tidal tails or a significant morphological disturbance corresponding to a major merger are detected in the optical image, we speculate that X2 could be the candidate for an activated IMBH, which has encountered a dense, CO-emitting molecular gas while wandering through the galactic disk \citep{seepaul22}, which was originally suggested by \citet{MH02} to interpret ULXs. There are several hypotheses about the origin of IMBHs. The IMBH could have formed within the galaxy due to a runaway merger of stellar black holes and stars \citep{MH02,PM02} or it could have been captured after a minor merger with a satellite or dwarf galaxy \citep{weller22}. The IMBH would then subsequently undergo dynamical relaxation into the galactic plane where most molecular gas is localized within giant molecular clouds. The expected mass range is $M_{\rm BH}\sim 10^2-10^5~\rm M_{\odot}$ \citep{GSH20}. The model of an IMBH traversing through a denser material is also supported by an increased column density of $N_{\rm H}\sim 2.3\times 10^{22}~{\rm cm^{-2}}$ indicating a mild obscuration, which is consistent with the molecular gas whose column density is expected to be $\sim 10^{21}-10^{22}~\rm cm^{-2}$ \citep{schneider15}, i.e. well below the Compton-thick limit of $\sim 1.5\times 10^{24}~\rm cm^{-2}$ that would cause a severe absorption of X-ray photons. The column density is consistent with the intrinsic obscuration of the IMBH embedded inside the giant molecular cloud (GMC) of $m_{\rm GMC}\sim 10^6\,M_{\odot}$ with the mean radius of $R_{\rm GMC}\sim 25\,{\rm pc}$~\footnote{The intrinsic column density then is $N_{\rm H}\sim 3 m_{\rm GMC}/(4 \mu m_{\rm H} \pi R_{\rm GMC}^2)\sim 2.4 \times 10^{22}\,(m_{\rm GMC}/10^6\,M_{\odot})(R_{\rm GMC}/25\,{\rm pc})^{-2}{\rm cm^{-2}}$, where $\mu\sim 2$ is the mean molecular weight of $\rm H_2$. }, which are typical parameters for a GMC \citep{HD15}. Since most of $\rm H_2$ gas is contained in GMCs, we expect about $M_{\rm H_2}/m_{\rm GMC}\sim 2500$ of them in HCG~97b.

In this regard, the energy and the momentum feedback from the accretion onto the X2 source is expected to contribute to the heating and the mechanical input into the surrounding ISM, which is then prone to a more enhanced ram-pressure stripping. This could be relevant since the footprint of the radio tail lies in projection close to the X2 source, see Figs.~\ref{fig:radio} and \ref{fig:radio_smo20} \footnote{Previously, the jellyfish galaxy JW100 was also reported to host a bright ULX source, however, it does not have any apparent connection with the observed radio tail \citep{poggianti19}}. This implies a novel mechanism, in which IMBHs or ULXs in general can strengthen the ISM stripping due to the IGrM ram pressure. Such a scenario is complementary to the studied case when ram-pressure stripping ignites an active galactic nucleus (AGN) activity \citep[see e.g.][]{peluso22}. \citet{poggianti17} proposed the scenario when an AGN can inject energy and momentum to enhance ram-pressure stripping, however, a potential connection with ULXs/IMBHs has not been studied. Also, currently the connection between the AGN activity and the ram-pressure stripping is unclear and some studies questioned the causal relation \citep[see e.g.][]{BFS22,cattotini23}, which further motivates us to study the impact of the ULX X2 source in HCG 97b. In addition, the acceleration of electrons due to the IMBH feedback and the associated shocks can help explain the flatter spectral index of the tail region. Below we provide several analytical estimates on how this can work for typical parameters of the IMBH, ISM densities and temperatures.    

Considering radiative efficiency of $\eta=0.1$ and the bolometric correction of $\kappa_{\rm bol}=13.3$, which is intermediate between low-luminosity and high-luminosity accretion modes (see Subsection~\ref{sec:bh}), we estimate the X2 accretion rate as follows,
\begin{align}
  \dot{M}&=\frac{L_{\rm bol}}{\eta c^2}=\frac{\kappa_{\rm bol}L_{\rm X}}{\eta c^2}\,\notag\\
  &\sim 4.22\times 10^{-5}\left(\frac{\kappa_{\rm bol}}{13.3} \right)\left(\frac{L_{\rm X}}{1.8\times 10^{40}\,{\rm erg\,s^{-1}}} \right)\left(\frac{\eta}{0.1}\right)^{-1}\,{\rm M_{\odot}\,yr^{-1}}. \label{eq_acc_rate}
\end{align}
where $L_{\rm X}=1.8\times 10^{40}~\rm erg~s^{-1}$ is the $2-10$~keV luminosity of the X2 source. The accretion rates of the order of $\dot{M}$ given by Equation~\eqref{eq_acc_rate} are consistent with the passages of IMBHs through molecular clouds as studied by \citet{seepaul22}, who predicted X-ray luminosities exceeding $10^{41}~{\rm erg~s^{-1}}$ for the brightest cases. Considering the whole range of IMBH masses, $M_{\rm BH}\sim 10^2-10^5\, \rm M_{\odot}$, the Eddington ratio is in the range of $\lambda_{\rm Edd}=\dot{M}/\dot{M}_{\rm Edd}\sim 0.02-19.1$, where the larger values correspond to smaller IMBH masses. Since the IMBH coincides with the CO emission peak and we assume that it is fueled by the accretion from the surrounding molecular gas, a quasispherical thick hot flow with a lower Eddington ratio is expected to develop \citep{seepaul22}. Such a Bondi-like flow has properties similar to ADAF and is also associated with powerful outflows that could have contributed to the radio-tail development via their mechanical feedback, or at least they increased the amount of gas escaping from the host galaxy in addition to the IGrM ram pressure.

Therefore, for the following estimates, assuming the ADAF and the IMBH mass range, we fix the Eddington ratio to $\lambda_{\rm Edd}=10^{-2}$, which implies the IMBH mass of $M_{\rm BH}\sim 3.5\times 10^5~ \rm M_{\odot}$. This set-up is also consistent with the assumed radiative efficiency of $\eta\sim 0.1$ \citep{YN14}.
To model the accretion from the CO-emitting gas, we consider the Bondi-Hoyle-Lyttleton-like accretion from the molecular cloud with the $H_2$ number density range of $n_{\rm MC}\sim 10^2-10^4\,{\rm cm^{-3}}$ and the ambient temperature of $T_{\rm MC}\sim 10-20\,{\rm K}$ \citep{seepaul22}, which fixes the ambient mass density $\rho$ and the ambient gas sound speed $c_{\rm s}$. Following the basic Bondi-Hoyle-Lyttleton accretion theory, the accretion rate at the outer radius $R_{\rm A}$ is given by \citep[see e.g.][and references therein]{seepaul22},
\begin{equation}
    \dot{M}_{\rm B}=\frac{4 \pi G^2 M_{\rm BH}^2\rho}{(v_{\rm rel}^2+c_{\rm s}^2)^{3/2}}\,,
    \label{eq_BHL_accretion}
\end{equation}
where $v_{\rm rel}$ is the relative velocity of the IMBH with respect to the local molecular gas. Due to the powerful outflows and conduction, the actual gas inflow rate through the IMBH horizon is scaled down with respect to $\dot{M}_{\rm B}$ following a simple power-law relation,
\begin{equation}
    \dot{M}_{\rm in}\sim \dot{M}_{\rm B}\left(\frac{R_{\rm in}}{R_{\rm A}} \right)^p\,,
    \label{eq_inflow_rate}
\end{equation}
where the effective inner radius of the flow is set to $R_{\rm in}=50 R_{\rm S}$ \citep{abramowicz02}, with $R_{\rm S}=2GM_{\rm BH}/c^2$ being the Schwarzschild radius. The outer radius is given by the gravitational capture radius $R_{\rm A}=2GM_{\rm BH}/(v_{\rm rel}^2+c_{\rm s}^2)$. The power-law slope in Equation~\eqref{eq_inflow_rate} is set to $p=0.5$, which is motivated by numerical calculations of hot flows \citep{YN14,ressler20}.

\begin{figure}
    \centering
    \includegraphics[scale=0.54]{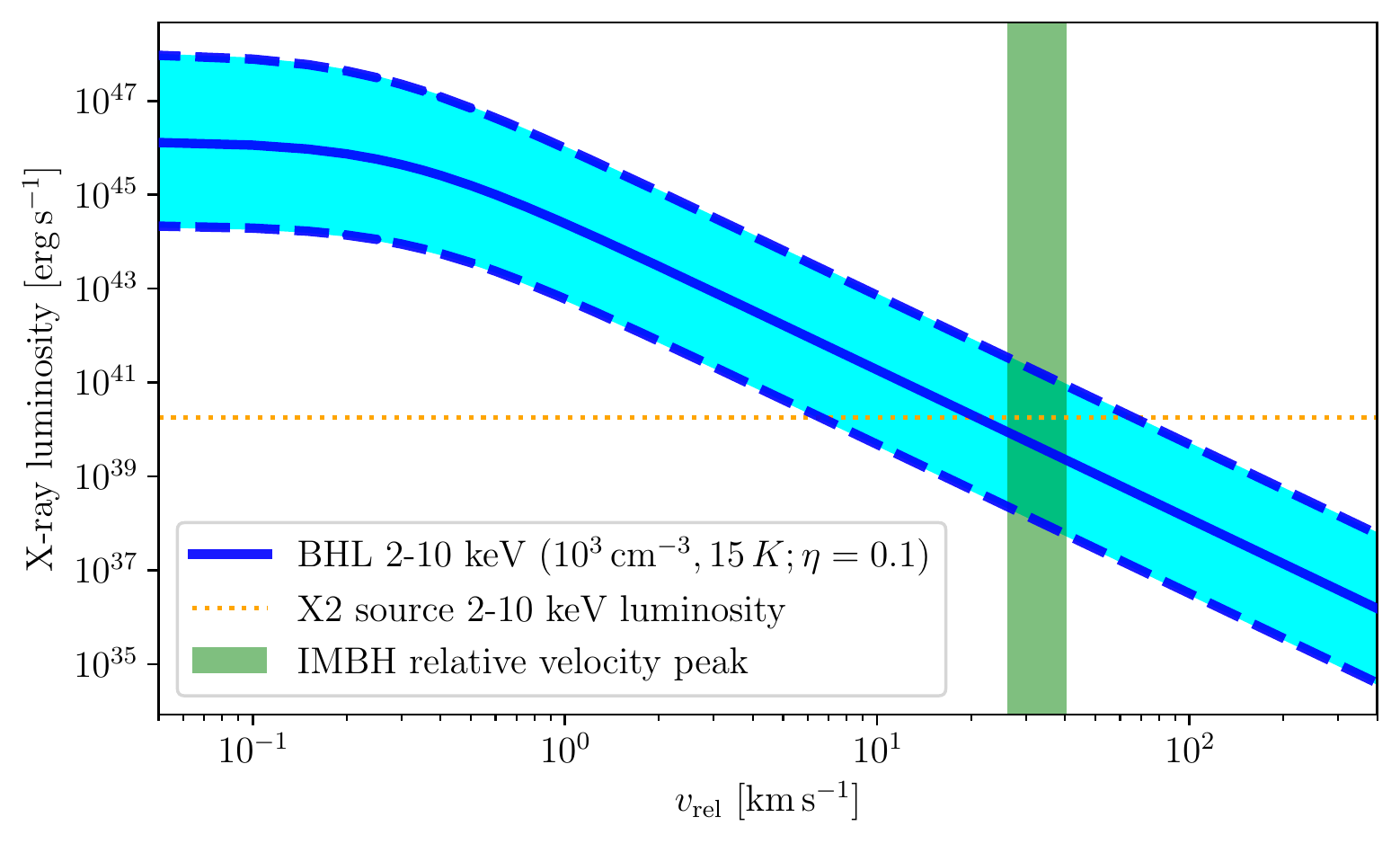}
    \caption{The $2-10$~keV luminosities corresponding to the Bondi-Hoyle-Lyttleton (BHL) accretion as a function of the relative velocity of a wandering IMBH with respect to the galactic disk molecular gas (solid blue line; the ambient molecular cloud has a number density of $10^3\,{\rm cm^{-3}}$ and temperature of $15\,{\rm K}$). The inferred X-ray luminosity of the X2 source is denoted by the orange dotted line. The cyan area and dashed blue lines express the BHL X-ray luminosities for the whole range of the molecular cloud characteristic parameters (number densities of $10^2-10^4\,{\rm cm^{-3}}$ and temperatures of $10-20\,{\rm K}$). The vertical green rectangle represents the estimated peak of the IMBH relative velocity distribution with respect to the local gas inferred from the Illustris TNG50 simulations \citep{seepaul22}.}
    \label{fig_BHL_accretion}
\end{figure}

To determine the required relative velocity $v_{\rm rel}$, we compare $\dot{M}_{\rm in}$, or rather the corresponding X-ray luminosity $L_{\rm X}\sim \eta \dot{M}_{\rm in}c^2/\kappa_{\rm bol}$ with the inferred 2-10 keV luminosity of X2. From Figure~\ref{fig_BHL_accretion}, we see that to reproduce the X2 X-ray luminosity, the IMBH is expected to move at the relative velocity of $\sim 6.6-66.3\,{\rm km\,s^{-1}}$ with respect to the local molecular gas, with $v_{\rm rel}\sim 21\,{\rm km\,s^{-1}}$ corresponding to the intermediate number density of $10^3\,{\rm cm^{-3}}$ and the temperature of $15\,{\rm K}$. This interval is consistent with the peak of the IMBH velocity distribution with respect to the surrounding gas inferred from Illustris TNG50 simulations \citep{weller22,seepaul22}, which is in the range of $26-41~\rm km~s^{-1}$. 

We further estimate if the activated IMBH of $M_{\rm BH}\sim 3.4\times 10^5~\rm M_{\odot}$ can significantly impact the surrounding molecular gas and induce the formation of the radio tail structure. The traversing IMBH crosses a typical molecular cloud in $\tau_{\rm cross}=D_{\rm MC}/v_{\rm rel}\sim 4.66 \times 10^5$ years for the molecular cloud diameter of $D_{\rm MC}\sim 10\,{\rm pc}$ \citep{HD15} and the relative velocity of $v_{\rm rel}\simeq 21\,{\rm km\,s^{-1}}$ (see the estimate above). During $\tau_{\rm cross}$, the ADAF flow surrounding the IMBH generates a significant outflow and potentially a relativistic jet that impacts the surrounding ISM via mechanical feedback as well as radiation pressure both from the accretion as well as the generated shocks when the outflow impacts the molecular cloud \citep[see e.g.][]{goicoechea16,FS16}. The observed 27-kpc radio tail of HCG~97b could have thus been formed and shaped via
\begin{itemize}
    \item[(i)] jet generated in the inner zone of the ADAF where the outflowing plasma is collimated by the ordered magnetic field and can reach ultrafast velocities in excess of $0.1 \rm c$ \citep[see e.g.][]{sukova21}, which is further distorted by the IGrM ram pressure, 
    \item[(ii)] uplift of the ISM material due to IMBH-induced mechanical feedback and subsequent IGrM ram-pressure prolongation and distortion.
\end{itemize}

The radio tail generation with the case (i) scenario can work if the inner ADAF surrounding the IMBH can support the ultrafast outflow/jet with the velocity of $v_{\rm UFO}\sim D_{\rm tail}/\tau_{\rm cross}\sim 0.2\,{\rm c}$ to be able to create the tail of the 27-kpc length during the crossing time. The restriction on the jet outflow velocity can be eased if the IMBH is fed by the molecular material for a longer duration than inferred from the typical cloud size. The scenario where radio jet is generated by an accreting IMBH is also supported by the recent detection of the two-sided jet ($L_{\rm 1.66~GHz} \sim 3\times 10^{38}$~$\rm erg~s^{-1}$) associated with the IMBH of $3.6\times 10^5\, \rm M_{\odot}$ in a dwarf galaxy \citep{yang23}. At 1.4 GHz, the tail luminosity of HCG~97b is comparable to that value, $L_{\rm 1.4~GHz}\sim 7\times 10^{37}\,{\rm erg\,s^{-1}}$ within an order of magnitude. The one-sided nature of the IMBH jet outflow, which could be associated with the north-western radio lobe visible at 4.86 GHz, can be caused by a small viewing angle and/or the distortion, bending, and further prolongation of the tail by the IGrM ram pressure. In this picture, the extended tail part could be the remnant of the previous accretion activity of the X2 as it passed through another denser molecular-cloud core. This scenario provides a possible explanation of the high velocity of the radio plasma in the radio tail, as discussed in Subsection~\ref{sec:rps-spcix}, and can be further studied by observing a small-scale radio emission associated with the X-ray point source X2. Furthermore, the flatter radio spectral index inferred for the tail can be interpreted as a result of the local electron acceleration process, such as in shocks created due to the interaction between the jet launched by the IMBH and the IGrM.

Case (ii) can be supported by comparing the mechanical energy density of the generated wide-angle outflow $\epsilon_{\rm out}$ with the thermal energy density of the surrounding gas $\epsilon_{\rm gas}\sim \rho c_{\rm s}^2$. The outflow rate for the ADAF $\dot{M}_{\rm out}$ is much larger than the inflow rate $\dot{M}_{\rm in}$ and can be expressed as $\dot{M}_{\rm out}=\dot{M}_{\rm B}[1-(R_{\rm in}/R_{\rm A})]^p$, which results in $\dot{M}_{\rm out}\sim 0.16\,{\rm M_{\odot}\,yr^{-1}}$ or about 2024 times more than the inflow rate for the intermediate molecular-cloud number density of $10^3\,{\rm cm^{-3}}$. The hot outflow velocity is estimated using $v_{\rm out}=0.2v_{\rm K}(R_{\rm A})\sim 3\,{\rm km\,s^{-1}}$ \citep{yuan18}, where $v_{\rm K}(R_{\rm A})\sim 14.8\,{\rm km\,s^{-1}}$ is the Keplerian velocity at $R_{\rm A}\sim 6.86\,{\rm pc}$. The hot outflow energy density can then be estimated as,
\begin{equation}
    \epsilon_{\rm out}\sim \frac{1}{2}\frac{\dot{M}_{\rm out}}{V_{\rm MC}}\tau_{\rm cross}v_{\rm out}^2=\frac{1}{2}\frac{\dot{M}_{\rm out}v_{\rm out}^2}{D_{\rm MC}^2v_{\rm rel}}\sim 2.3\times 10^{-10} {\rm erg\,cm^{-3}}\,\label{eq_outflow}
\end{equation}
where $V_{\rm MC}\sim D_{\rm MC}^3$ is the molecular cloud volume. We compare $\epsilon_{\rm out}$ with the thermal energy density of the molecular gas, $\epsilon_{\rm gas}\sim n_{\rm gas}k_{\rm B}T_{\rm gas}\sim 2.1 \times 10^{-12}\,{\rm erg\,cm^{-3}}$, as well as with that of the radio-emitting warm and hot ionized plasma, $\epsilon_{\rm gas}\sim (0.3\,{\rm cm}^{-3})\times (1.38\times 10^{-16}\,{\rm erg\,K^{-1}})\times 9000\,{\rm K}\sim 3.7\times 10^{-13}\,{\rm erg\,cm^{-3}}$ and $\epsilon_{\rm gas}\sim (3 \times 10^{-3}\,{\rm cm^{-3}})\times (1.38\times 10^{-16}\,{\rm erg\,K^{-1}}) \times (5\times 10^6\,{\rm K})\sim 2.1\times 10^{-12}\,{\rm erg\,cm^{-3}}$, respectively \citep[see e.g.][]{seepaul22}, and we see that $\epsilon_{\rm out}$ is larger by two-three orders of magnitude. In order for the gas to escape from the galaxy, $\epsilon_{\rm out}$ needs to be larger than the gravitational binding energy density $\epsilon_{\rm bind}$. For a more diluted ionized gas in the warm-hot interstellar medium, we estimate $\epsilon_{\rm bind}$ assuming the virial theorem and using the mean rotational velocity $v_{\rm rot}\sim 218\,{\rm km\,s^{-1}}$ (see Subsection~\ref{sect:CO}),
\begin{align}
\epsilon_{\rm bind}&\sim \frac{1}{2}\frac{M_{\rm gas}}{V_{\rm gas}}v_{\rm rot}^2\sim \frac{1}{2}\rho_{\rm gas}v_{\rm rot}^2\,\notag\\
&\sim 4\times 10^{-11} \left(\frac{n_{\rm gas}}{0.1\,{\rm cm^{-3}}} \right) \left(\frac{v_{\rm rot}}{218\,{\rm km\,s^{-1}}} \right)^2\,{\rm erg\,cm^{-3}}\,,\label{eq_bind}
\end{align}
which is smaller than $\epsilon_{\rm out}$ for typical warm/hot ionized medium number densities.
Overall, this would lead to the uplift of  gas and its subsequent easier stripping by the ram pressure of the IGrM. Currently, only the radio tail tracing the ionized plasma is observed. Future more sensitive observations, e.g.  H$\alpha$+[NII] narrow band imaging, will constrain the amount of the stripped neutral gas. However, the mechanical uplifting of the galactic ISM gas works more efficiently for the less dense ionized component, which follows from the comparison of Eqs.~\eqref{eq_bind} and \eqref{eq_outflow}. This can explain the lack of the neutral and molecular gas in the ram-pressure stripped tail since it rather stays within the galactic disk due to larger binding energy.

The lower-Eddington ratio ADAF model for the X2 X-ray emission is motivated by the plausible accretion from the ISM \citep{seepaul22}. The Bondi-like hot flow is also associated with a nonthermal radio emission due to the ADAF spectral energy distribution \citep{YN14}. Moreover, a wide-angle outflow is naturally generated; hence, mechanical feedback could contribute to the detected stripped material. 

If instead, the accretion of the ambient dense gas proceeds with an Eddington ratio close to unity, and hence in the radiatively efficient mode, the X2 mass would be pushed to lower values, though still in the IMBH mass range,
\begin{equation}
    M_{\rm BH}\simeq 300.4 \left(\frac{L_{\rm X}}{1.8\times 10^{40}\,{\rm erg\,s^{-1}}} \right)^{1.3} \left(\frac{\lambda_{\rm Edd}}{1} \right)^{-1}\,M_{\odot}\,,
\end{equation}
which follows from Equation~\eqref{eq_BHmass}, where we express the bolometric correction $\kappa_{\rm bol}$ following \citet{netzer19} ($\kappa_{\rm bol}\sim 2.1$ for the X2 $2-10$~keV luminosity) and hence the standard disk formalism. We note that the Eddington limit would have to be exceeded by a factor of at least 3 for the X2 to reach stellar-mass black hole masses and the neutron-star mass range would be reached for $\lambda_{\rm Edd}\sim 200$ (see Table~\ref{tab:bh_mass} for the BH mass-range estimates for the Eddington ratio between $10^{-2}$ and $10^2$). For $\lambda_{\rm Edd} \sim 1$, the accretion flow has the structure of a geometrically and optically thick slim disk \citep{abramowicz88} or a ``puffy'' disk \citep{lancova19}, for which radiation-driven outflows are expected \citep{feng19}. 
Since the outflow is radiatively driven, we can estimate the radiative feedback to the surrounding medium from the bolometric accretion luminosity, assuming the efficient mode during the whole passage of the IMBH through the molecular cloud,
\begin{equation}
    \epsilon_{\rm rad}\sim \frac{\kappa_{\rm bol}L_{\rm X}\tau_{\rm cross}}{D_{\rm MC}^3}=\frac{\kappa_{\rm bol}L_{\rm X}}{D_{\rm MC}^2 v_{\rm rel}}\sim 1.9\times 10^{-5}\,{\rm erg\,cm^{-3}}\,,
\end{equation}
where we assumed that the released radiative energy during $\tau_{\rm cross}$ is primarily impacted within the molecular cloud volume. We see that $\epsilon_{\rm rad}$ is larger than typical gas thermal energy and binding energy densities $\epsilon_{\rm gas}$ and $\epsilon_{\rm bin}$ estimated previously by several orders of magnitude, hence the radiative feedback expected from a lower-mass IMBH or a stellar-mass BH is expected to have profound effects on the surrounding molecular gas, which should result in the severe disturbance and uplifting of bound gas. In this case, we would expect a harder X-ray emission due to a smaller black-hole mass, which would result in the ionization of the neutral and molecular gas that is observationally not detected in the tail of HCG~97b. 

More constraints on the X2 mass will follow from the future detailed X-ray spectroscopy and timing, which will also help clarify the accretion mode and the type of feedback. In addition, magnetohydrodynamical simulations may be necessary to fully capture the interaction between the IMBH-generated outflow and the ISM medium, as well as the IGrM ram pressure, which is beyond the scope of the current study.   

Given that HCG~97b is slightly larger than the Milky Way, both X2 and X1 X-ray sources are located relatively in the central regions of HCG~97b, i.e. within $\sim 5.2\,{\rm kpc}$ from the nucleus, where the occurrences of both the molecular gas and IMBHs are statistically increased, and hence also the probability of their interaction \citep{HD15,weller22,seepaul22}. The X1 source could also be the candidate for an IMBH with $\sim 0.86$~kpc offset from the nucleus, though its location does not coincide with the CO emission peak, which is supported by the smaller obscuration in comparison with X2. Therefore X1 is not a prime candidate for a wandering IMBH reactivated by the interaction with the molecular gas \citep{seepaul22}. 
The constrained steeper power-law index ($\Gamma=2.65\pm 0.87$; Subsection~\ref{sect:xray}) of X1 indicates a soft state and a higher Eddington ratio, which makes it a candidate for a stellar-mass ULX with the mass of $\sim 39.5~\rm M_{\odot}$ for the Eddington limit.

\section{Conclusions}
In this paper, we present a multi-wavelength analysis of the spiral galaxy HCG~97b, and the main results are summarized as follows:
\begin{enumerate}
    \item LOFAR 144~MHz, VLA 1.4~GHz and 4.86~GHz images reveal asymmetric, extended radio continuum emission of the spiral galaxy HCG~97b, including an elongated radio emission along the optical disk and an extraplanar, one-sided radio tail extending $\sim 27$~kpc toward the group centre at GHz frequencies and to $\sim 60$~kpc at 144 MHz. 
    \item Chandra images reveal two off-centre X-ray sources with $2-10$~keV luminosities of $\sim 10^{39} - 10^{40}$~$\rm erg~s^{-1}$. Given the off-nuclear location and the observed X-ray luminosity, these two X-ray sources are most likely ULXs. 
    \item The source X2 is a suitable candidate for an accreting IMBH embedded in an environment with an increased density of molecular gas. 
    \item Asymmetry in the molecular gas within the disk is evident from both CO emission morphology and kinematics, indicating that HCG~97 is experiencing ram-pressure stripping, with the leading side at the southeastern edge of the disk. 
    \item Based on estimations of the capabilities of gravitational and hydrodynamic processes, it seems HCG 97b has experienced gravitational interaction and is currently subjected to ram pressure stripping. Both mechanisms have played roles in gas stripping, with the gravitational interaction potentially flattening the galaxy's gravitational potential well, making the gaseous components of the galaxy more susceptible to being stripped away by ram pressure.
    \item The VLA 4.86 GHz image reveals two bright radio blobs --- one above the disk and another outside of it, aligning with the radio tail. Their proximity to source X2, along with comparable and flat spectra indices ($\alpha > -1$) in the disk and tail, suggests that these blobs may be a pair of radio lobes recently powered by ULX feedback. 
    \item Considering the radiative time of CRe in the tail is $2.2\times 10^7$~yr, the expected velocity needed for transporting CRe from the disk to the tail is approximately 1300~$\rm km~s^{-1}$. This velocity substantially exceeds those observed in other ram-pressure stripped galaxies ($100-600$~$\rm km~s^{-1}$; \citealt{ignesti23}), and aligns with velocities found in head-tail radio galaxies \citep{ignesti20,edler22}. Therefore, the formation of the radio tail might also be influenced by the putative IMBH-induced activity.
\end{enumerate}

\section*{Acknowledgements}

We sincerely thank the referee for providing valuable comments and suggestions.
We thank Chentao Yang, Junjie Mao, Stefan William Duchesne, and Muryel Guolo for their valuable assistance and helpful discussion.
This research was supported by the GACR grant 21-13491X.
RG acknowledges support from the project LM2018106 of the Ministry of Education, Youth and Sports of the Czech Republic and project RVO:67985815.
AI acknowledges funding from the European Research Council (ERC) under the European Union’s Horizon 2020 research and innovation programme (grant agreement No. 833824 and the INAF founding program `Ricerca Fondamentale 2022' (PI A. Ignesti).

We acknowledge the use of the Legacy Surveys data and the Helpdesk for guidance in generating a composite DECaLS image.

\section*{Data Availability}

The data in this article are available on request to the corresponding author.


\bibliographystyle{mnras}
\bibliography{references} 




\bsp	
\label{lastpage}
\end{document}